%% file: ms.tex
\documentclass[useAMS,usenatbib]{mn2e} 
\usepackage{comment}
\usepackage{ifthen} 
\usepackage{amsmath} 
\usepackage{amsfonts}
\usepackage{amssymb} 
\usepackage{deluxetable} 
\usepackage{graphicx}
\usepackage{wasysym} 
\usepackage[caption=false]{subfig}
\usepackage{lscape}

\input{newcommand_gen.tex}

\title[150 New transiting planet candidates
from {\em Kepler} Q1-Q6 data]
{150 New transiting planet candidates from {\em Kepler} Q1-Q6 data}

\author[X.Huang, G.\'A.Bakos, J.D.Hartman]{X.Huang$^{1}$\thanks{E-mail:xuhuang@princeton.edu;},
G.Bakos$^{1}$,
J.D.Hartman$^{1}$,\\
$^{1}$ Department of Astrophysical Sciences, 4 Ivy Lane, Peyton Hall, 
Princeton University, Princeton, NJ 08544}

\begin{document}

\maketitle

\begin{abstract} 
We have performed an extensive search for planet candidates
in the publicly available {\em Kepler} Long Cadence data from quarters 
Q1 through Q6. The search method consists of initial de-trending of the 
data, applying the trend filtering algorithm, searching for transit 
signals with the Box Least Squares fitting method in three frequency 
domains, visual inspection of the potential transit candidates, and 
in-depth analysis of the shortlisted candidates. 
In this paper we present 150 new periodic planet candidates and 7 
single transit events, 72 of which are in multiple systems. The 
periods of these planet candidates vary from $\sim$0.17\,day to 
$\sim$440\,day. 124 of the planet candidates have radii smaller than 
3\,\rearth. We recover 82.5\% of the \citet{Batalha:2012} KOI catalog. 
We also report 40 newly identified false positives---systems that look 
like transiting planets, but are probably due to blended eclipsing 
binaries. Our search improves the statistics in the short period and 
small planet radii parameter ranges. 
\end{abstract}

\begin{keywords}
planetary systems-stars: individual 
\end{keywords}

\section{Introduction}
\label{sec:intro} 

The field of transiting extrasolar planets (TEPs) has exploded over the
past few years. One major contributor is the {\em Kepler Space Mission},
continuously monitoring 156000 stars in a $115\,deg^2$ field 
\citep{Borucki:2010,Koch:2010}. 
Since its launch in 2009, it has found 2321 planetary transit candidates\footnote{http://kepler.nasa.gov/Mission/discoveries/candidates/}
\citep{Borucki:2011a,Batalha:2012}. 
The first 4 months of data (Q1 and Q2) yielded 1235 planet candidates 
associated with 997 host stars, including 60 confirmed planets around 
33 stars \citep[hereafter B11]{Borucki:2011b}.
The {\em Kepler} team has developed increasingly sophisticated 
procedures to identify planet candidates \citep{Smith:2012} and 
multiple systems \citep{Steffen:2012,Ford:2012}. Using more data (16 
months) and improved detection procedures \citep[the Transit Planet Search 
(TPS) algorithm;][]{Jenkins:2010c, Tenenbaum:2012}, the total number of planet 
candidates has almost doubled since the release by B11 
\citep[hereafter B12]{Batalha:2012}. 

An independent search with different tools can build confidence in the 
reliability of the detections, and the completeness of the sample, by 
e.g.~providing new candidates that were missed by the {\em Kepler} 
science team. One such effort is the citizen science initiative, called 
PlanetHunters \citep{Fischer:2012,Lintott:2012}, based on the idea of 
Zooniverse \citep{Lintott:2008}. Making use of human eyes to search for
transit-like events through a user friendly computer interface, 
6 new planet candidates were published by far 
\citep{Fischer:2012, Lintott:2012}. These were then subjected to the vetting 
procedure of the {\em Kepler} team, and five of them survived this 
process, i.e.~they are not false alarms, as much as it can be determined 
from the {\em Kepler} data. However, conducting a visual search of all the 
raw \lcs\ takes a lot of human effort. Quoted from \citet{Fischer:2012}: 
``It is impractical for a single individual to review each of the 
$\sim150,000$ \lcs\ in every quarterly release of the {\em Kepler} data 
base.'' While challenging, it is, however, feasible for an individual to 
examine $\sim150,000$ \lcs, if sophisticated computer algorithms narrow 
down the list to a somewhat smaller sample of candidate transits that 
can be then checked very carefully. Also, note that small planets are 
often hard to recover by visual inspection without phase-folding the 
light curve at the suspected periodicity of the signal.

Encouraged by these observations, we started an independent search for
transiting planet candidates in {\em Kepler}'s long-cadence data. 
{\em Kepler} observations are grouped into so-called quarters (each 
having a duration of 3 months, except for Q0 and Q1, which are 
shorter). At the end of each quarter, the spacecraft rotates 90 
degrees, to adjust its solar panels.
The majority of {\em Kepler} stars are observed as long cadence (LC) 
targets, for which data are obtained by gathering over 270 exposures 
within 29.4\,min; 512 stars are also observed in short cadence (SC) 
mode with 58.9\,s intervals 
\citep{Jenkins:2010b,Gilliland:2010,Murphy:2012}. We used quarters 
Q0-Q6 in this search, which data were released to the public in Jan 2012. 

Our methodology is based on our experience conducting a similar search 
with HATNet \citep{Bakos:2004}, a wide-field ground-based survey. 
Broadly speaking, the {\em Kepler} space-based data is of much higher
quality than any ground-based data. Ground based observations often 
exhibit large gaps in the time-series, either due to the rotation of 
the Earth, or inclement weather conditions. The data quality is highly 
variable due to changing extinction, clouds, background, seeing, etc. 
The per-point photometric precision is typically worse than from space, 
partly because of the above effects of ground-based observations, and 
partly due to the use of inexpensive hardware (e.g.~front illuminated 
CCDs). Altogether, our ground-based data is of lower signal-to-noise, 
has inhomogeneous quality, and exhibits complex systematic variations 
with long gaps. For example, not a single transit event has been 
found in HATNet data by direct visual inspection; transits are 
detected through sophisticated data mining and phase-folding. 
Also, no robust {\em single} transit event was ever found by HATNet.
Consequently, tools developed for a transit search using ground based 
data may be very efficient in recovering transit signals from 
{\em Kepler}.

In this paper we employ tools from the transit detection pipeline of 
the HATNet project, after sufficient modifications to conform to the 
{\em Kepler} data. We also develop a pre-filtering method that 
corrects the {\em Kepler} \lcs\ for known anomalies and systematics, 
before searching them for transit events. Our search is blind in 
the sense that the list of {\em Kepler} candidates was not consulted 
during the search.   

The structure of the paper is constructed as follows. The data 
processing methodology is discussed in \S \ref{sec:Method}. We laid out 
our findings in \S \ref{sec:Result}, including the recovery of Kepler 
planet catalog, our selection and modeling of new candidates, and the 
properties of these new candidates. We make our concluding remarks 
in \S \ref{sec:Conc}.

\section{Data analysis}
\label{sec:Method}

%
\subsection{Removal of points and long trend filtering} 
\label{sec:long}

We make use of the {\em Kepler} public LC \lcs.
We begin with the TIME and SAP$\_$FLUX (raw flux) columns in the FITS 
files. The raw flux is already corrected for cosmic rays and background 
variations by the {\em Kepler} team \citep{Jenkins:2010a}. First we 
convert the fluxes to magnitudes and set the mean value for each star 
to its {\em Kepler} magnitude taken from the Kepler Input Catalog 
(KIC)\footnote{http://tdc-www.harvard.edu/software/catalogs/kic.html}.

The second step is to clean the \lcs\ based on the data anomalies table 
in the {\em Kepler Data Release Notes} for each quarter
\citep{Christiansen:2012,Christiansen:2012b,Machalek:2010,Machalek:2011}.
We summarize all the important events in Q1 through Q6 in Table 
\ref{tab:anomaly}. 
We also describe the definition of anomaly types (adopted from 
{\em {Kepler} Data Characteristic Hand Book}) and our methods of 
correction for each type in the table notations. 
Generally, for data anomalies involving a discontinuity we model the 
jump by a polynomial with an offset in magnitude after the gap. The 
fitting uses 50 points on both sides around the gap. The offset is then 
subtracted from the data after the jump.
In the time range during safe modes (Q2 and Q4), or earth point 
recoveries (Q3, Q4 and Q6), there is an exponential decay in the flux. 
We identify and remove the whole exponential decay rather 
than attempting to correct for the effect. 

Most of the stars exhibit long-term trends. We follow the traditional 
high-pass filter method \citep{Ahmed:1974,Mazeh:2010}, and apply a 
cosine filter on all the cleaned \lcs. For each \lc, before computing 
the filter, we generate a model by applying a 100 point ($\sim 2$ day 
long) median filter. This is done to prevent distortions due to outlier 
points (including introducing spurious ``transit'' signals).
The cosine filter is then computed as the sum of a linear component and 
$N=T_{total}/\Delta{T}$ cosine functions, where the highest frequency 
is $1/\Delta{T}$ ($\Delta{T}=1\,\mathrm{day}$),
and $T_{total}$ is the total time span of the \lc:
\begin{equation} 
M(t_j)=a\frac{(t_j-t_s)}{T_{total}}+\sum_{i=0,N}{b_i\cos\left[i\pi\frac{(t_j-t_s)}{T_{total}}\right]}.
\end{equation} 
Here $t_j$ is the time of the $j$th measurement, and $t_s$ is the first 
time instance in the \lc. The coefficient $a$ for the linear component 
and coefficient $b_i$ for the $i$th cosine functions are computed by a 
least squares fitting procedure on the model. The fitted trend $M(t_j)$ 
is then subtracted from the \lcs. 
We apply this cosine filter to \lcs \ in every Quarter separately, 
and then combine the long trend filtered \lcs\ from Q1-Q6 by offsetting
the magnitude of all the quarters referring to the magnitude of Q1.

%
\subsection{Sky groups and Systematic Trend filtering}
\label{sec:TFA}

Following the long-term trend filtering procedure described above
(\refsec{long}), we then apply the Trend Filtering Algorithm (TFA) 
developed by \citet{Kovacs:2005} on the combined (Q1--Q6) \lcs. 
The idea of TFA is to select a set of template \lcs, which we assume to 
contain information of the systematic variations, and then construct a 
linear filter based on their shared time-series for each \lc\ to be 
corrected. TFA can remove systematic variations that are either shorter 
time-scale than those corrected by the cosine filter, or have an arbitrary 
functional form that is not well described by the sum of cosine 
functions. TFA assumes that the \lcs\ are sampled at the same 
time-instances.

To construct a set of template \lcs\ with the same time-base, we make 
use of the sky group information provided by the {\em Kepler} team. 
The sky group number is defined as the CCD channel number on which the 
stars fell during Q2 of the {\em Kepler} operation.  
{\em Kepler} has 21 modules. Each module contains 2 CCD chips and each 
CCD contains 2 output channels. The focal plain rotates 90 degrees when 
the telescope switches to a new quarter every 3 months, except for the 
initial transition between quarters Q0 to Q1. Generally, the stars that 
belong to the same sky group share the same CCD channel all the time, 
although this channel changes from time to time. Therefore, each sky group 
shares the same time base, and has similar instrumental systematic trends. 
Additionally, stars in the same sky group are related in terms of their 
sky position, so that the local variabilities (e.g.~local background 
variations or flux contamination from nearby stars) are often shared. 
TFA is designed to reduce this kind of general (shared by a number of 
stars) systematics. 

We construct a separate filter for each sky group, selecting
$\sim300$ template \lcs\ in each case. Template stars are selected 
randomly, but in a manner that ensures a uniform distribution of their 
positions across the field. To exclude variable stars, we also impose a 
constraint on their median deviation around the median magnitudes 
(MAD, a quantity that is insensitive to outliers); stars with high MAD 
are ruled out from the templates. The total number of data points 
(time samples) in each template time base is $\sim\,17000-22000$, 
depending on the total length of time series in the sky group. In other 
words, we are not overfitting the \lcs.

Altogether 124840 stars in 84 sky groups were selected and analyzed 
with TFA from the {\em Kepler} public LC data. For each sky group, 
we only selected stars that have been observed during the complete 
time range, and have not been affected by the failure of module 3. 
This is because the TFA analysis requires the same time-base to 
generate the filter per sky group. Module 3 failed in the 
middle of Q4, while observing sky groups 5, 6, 7 and 8. Due to the 
rotation of spacecraft, the Q5 data in sky groups 49, 50, 51 and 52, 
and the Q6 data in sky groups 77, 78, 79, and 80 were not available. 
Stars in these sky groups are still included if they have a complete 
data set in other quarters. Otherwise, we only selected \lcs\ 
containing all the Q1-Q6 data (which has a total observation time of 
$\sim\,$500\,days).

We note that the {\em Kepler} team has recently implemented a new
``cotrending'' algorithm, called PDC-MAP \citep{Smith:2012}. This
algorithm uses 16 Cotrending Basis Vectors (CBV) that are generated by
a singular value decomposition method applied separately to each
channel and each quarter. This method has some commonalities with
TFA. We did not make use of the PDC-MAP data, since it was not
available for the Q1--Q6 data when we started our analysis.

\begin{figure*} 
\includegraphics[angle=0,width=0.9\textwidth]{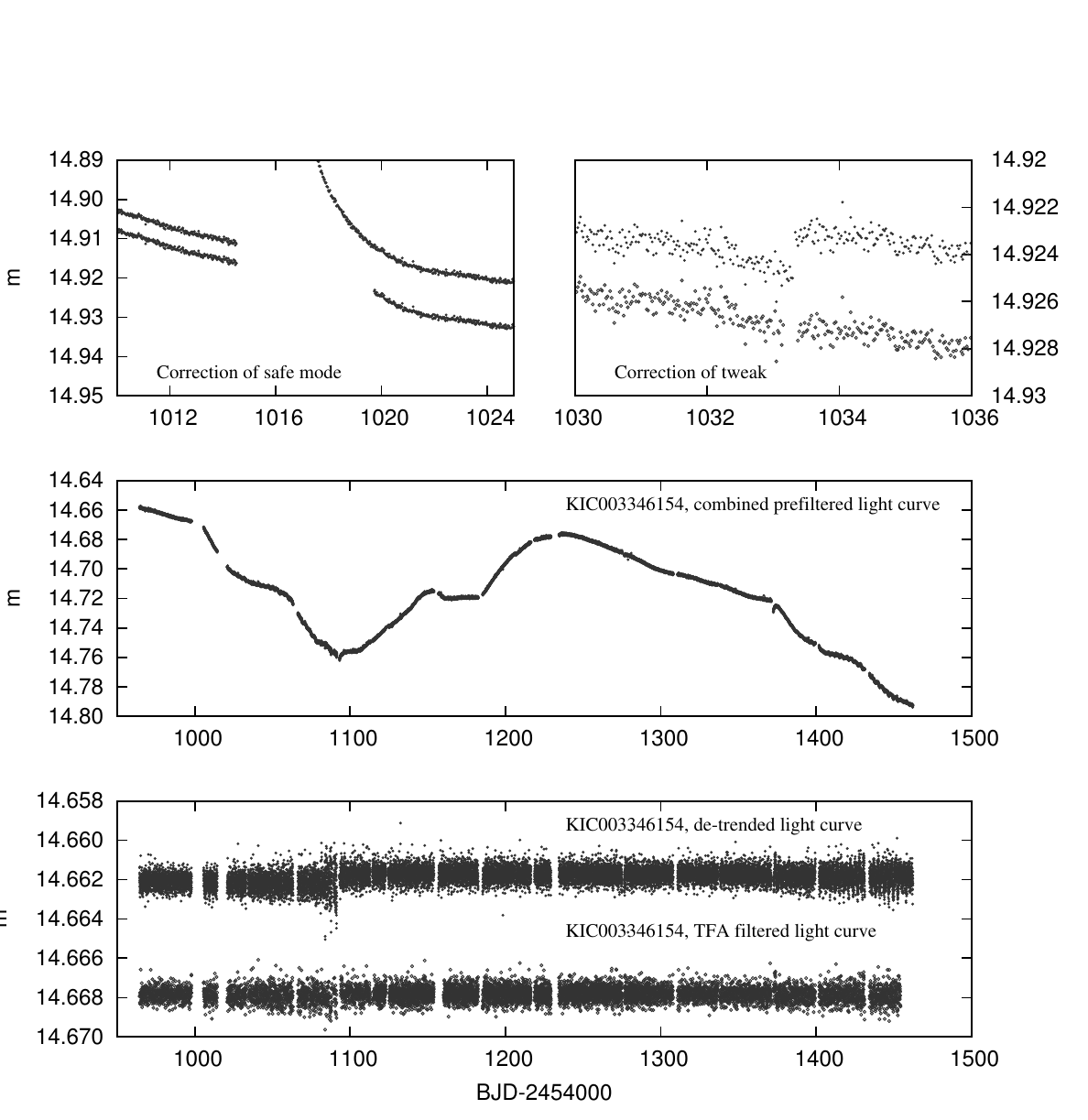}
\caption{
An example of our filtering process as applied to the \lc\ of KIC 003346154. 
The  
ordinate is in the units of approximate {\em Kepler} magnitude: 
Top panel: correction of safe mode (left) and tweak (right), 
(the example is taken from Q2). The corrected \lc\ is plotted below the 
uncorrected one. 
Middle panel: combined \lc\ from Q1-Q6 after the pre-filtering process, 
i.e. correction of anomalous data points and combining the \lcs\ from all 
the quarters.
Bottom panel: \lc\ after long trend filtering (solid symbols) and after
TFA (empty circles).
\label{fig:filter}
}
\end{figure*}

In Figure~\ref{fig:filter} we demonstrate our filtering process on
a randomly selected \lc\ (KIC\,003346154).
The great improvement by the procedure is clearly demonstrated.

%
\subsection{Box Least Square Fitting (BLS) and Transit Analysis}

We use the box least square fitting algorithm \citep{Kovacs:2002} 
to search for periodic transit signals in the TFA-filtered \lcs. 
In order to maximize the efficiency of the BLS search in a wide 
frequency range, we divide the frequency search into three, only slightly 
overlapping frequency domains: 
$0.3\,\mathrm{d}^{-1}\,<f_1<\,9.0\,\mathrm{d}^{-1}\,$, 
$0.02\,\mathrm{d}^{-1}\,<f_2<\,1.0\,\mathrm{d}^{-1}\,$ and 
$0.005\,\mathrm{d}^{-1}\,<f_3<\,0.03\,\mathrm{d}^{-1}$. 
We use a different number of frequency steps and BLS bins in each 
domain. We use the SNR (signal to noise ratio, measured in the BLS 
spectrum) and DSP (dip significance) parameters 
\citep[for details, see][]{Kovacs:2002} of the first five frequency 
peaks reported by BLS to select candidate transit signals for manual 
inspection. 

We adopt selection threshold of 11 and 8.5 for SNR and DSP in the middle 
and long period range. The selection threshold of short period range is 
somewhat higher (30 and 20 for SNR and DSP) considering that BLS 
tends to respond to short period easier and there are fewer KOIs in this 
period range as our reference. On top of these requirements, we neglect the 
BLS peaks with frequency too close to the frequency domain boundary (i.e.
the period ranges used in selection are  
$0.67\,\mathrm{d}^{-1}\,<f_1'<\,6.67\,\mathrm{d}^{-1}\,$, 
$0.0202\,\mathrm{d}^{-1}\,<f_2'<\,0.67\,\mathrm{d}^{-1}\,$ and 
$0.005\,\mathrm{d}^{-1}\,<f_3'<\,0.0202\,\mathrm{d}^{-1}$). We also reject 
those with a transit duration much longer than expected 
($q_1<0.5$, $q_2<0.1$ and $q_3<0.013$) or a transit depth indicating 
very large planetary radius ($dip>0.04$). These are not optimized selection 
criteria, but instead ensure that no shallow or rare transit events are 
missed. The low detection threshold also means that events will require 
close visual inspection. With the above limits, we selected $\sim$10\% of 
the stars to fold with the BLS peaks that satisfy the above requirements. 
The number of BLS peaks selected is $\sim$3\% of all the first five best 
BLS peaks. We then manually inspected all the folded \lcs.

\begin{figure}
\includegraphics[angle=0,width=0.9\linewidth]{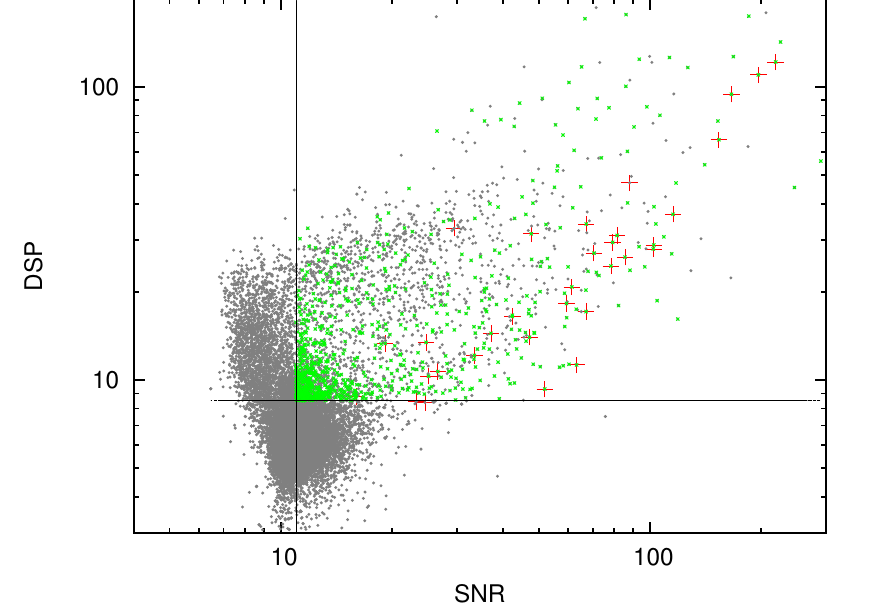} 
\caption{
A demonstration of our selection methods. SNR versus DSP for all the first 
five BLS peaks from stars in sky groups 1, 12, and 19 (black dots). Solid lines 
present the lowest limit of SNR and DSP for selection. Green asteroids 
resemble all the BLS peaks actually selected for folding considering the 
period range and transit duration. Red crosses show the most significant peak 
of all the KOI planet candidates (34 in total) in these sky groups.   
\label{fig:SNRDSP}
}
\end{figure}

We take three random sky groups as an example to illustrate our 
selection process in Figure \ref{fig:SNRDSP}. We show the distribution of 
BLS peaks (the mid period range) in the SNR-DSP plane. The solid lines are 
our lower limits for selection. The green dots are selected periods for 
further examination. The best BLS peak of KOI planet candidates are 
represented with red crosses. Majority of the KOIs are selected with 
high SNR and DSP.

Transit-like features in the folded \lcs\ are flagged and further 
examined in the visual inspection. We reject \lcs\ with recognizable depth 
variation for odd and even peaks. We also check the harmonics of the 
detected period to ensure the detection of a correct period. For the 
transits visible in the unfolded \lcs, we directly inspect the shape and 
transit center of each transit; for the transits invisible in the 
unfolded \lcs\ (due to low S/N), we perform phase-folding with the 
BLS-detected frequency before examining the data.

\section{Results} 
\label{sec:Result}

%
\subsection{Comparison of Our Sample to that of the Kepler Team}
\label{sec:comparKOI}
We flagged 2180 stars as possible transiting planet hosts during 
manual inspection. Among them, we found 180 stars which are categorized 
as eclipsing binaries (EB) 
\footnote{http://archive.stsci.edu/kepler/eclipsing$\_$binaries.html}
\citep{Prsa:2011} or false positives (FP)
\footnote{http://archive.stsci.edu/kepler/false$\_$positives.html}
\citep{Borucki:2011b} by the {\em Kepler} team.
We also cross-matched our results with the KOI catalog
\footnote{http://archive.stsci.edu/kepler/planet$\_$candidates.html} 
\citep{Borucki:2011a,Borucki:2011b,Batalha:2012}.
We note that our detection efficiency is comparable to the TPS algorithm 
used by the Kepler team \citep[Hereafter T12]{Tenenbaum:2012}. Their 
algorithm yields 5392 detections and detected 88.1\% of the 1235 KOIs 
in the B11 catalog using data from Q1-Q3. We recovered 92\% of the B11 
catalog in our analysis (1124 of the KOIs in B11 are in our initial data 
set,from which, 1034 are flagged) thanks to the longer time base we used 
compared to T12.  

\begin{figure}
\includegraphics[angle=0,width=0.9\linewidth]{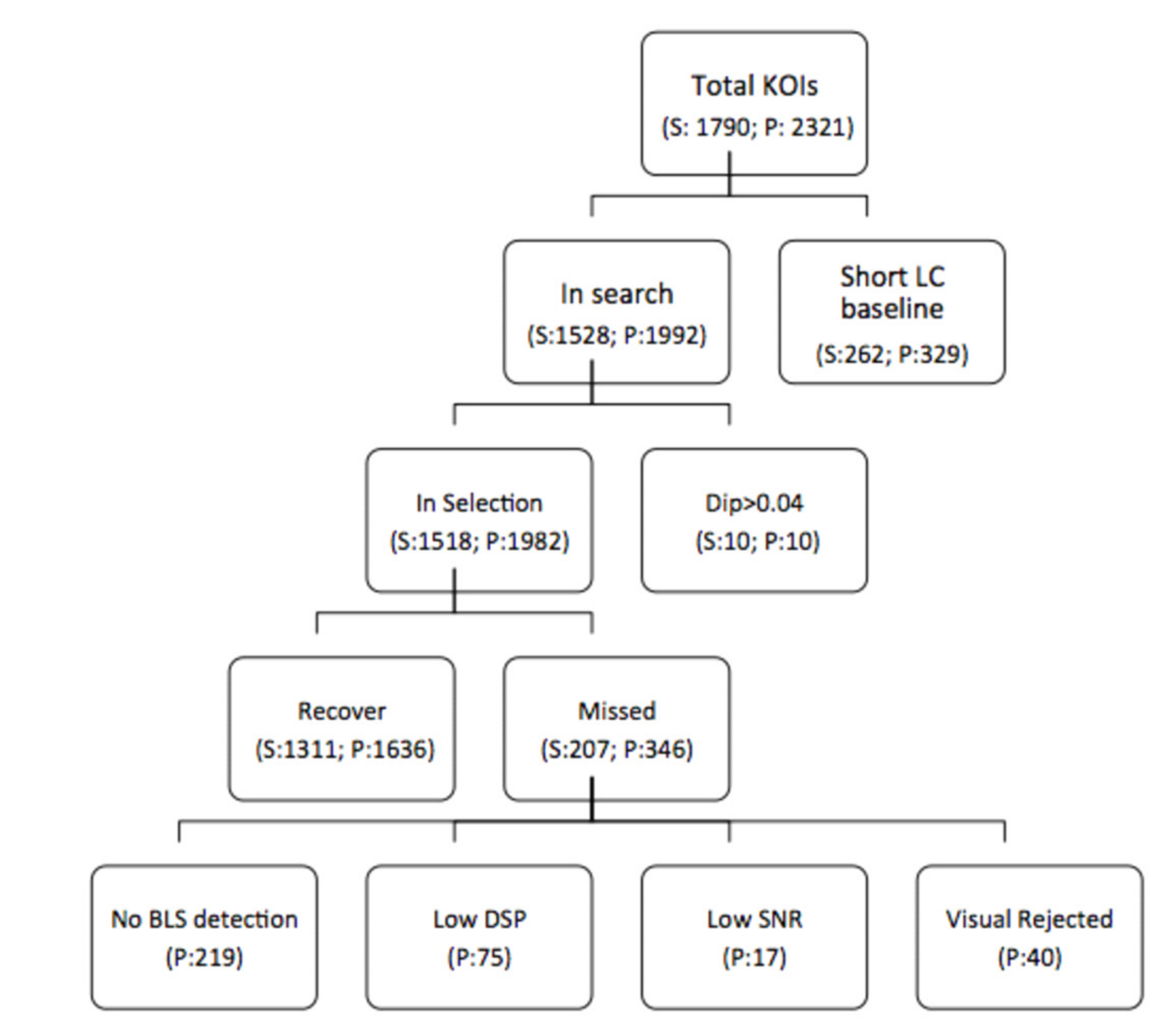} 
\caption{
Summary of our recovery of KOIs in the B12 catalog. We show
the number of stars (S) and planet candidates (P) in each category. 
The recovery rate of KOI host stars in our search range is 86.4\%. The
recovery rate of KOI planet candidates is 82.5\%. For more information,
see \S \ref{sec:comparKOI} and Table \ref{tab:KOItable}. 
\label{fig:summary}
}
\end{figure}

We demonstrate the selection of KOIs compared to the B12 catalog in 
Figure \ref{fig:summary}. There are 1518 KOI stars (corresponding to 1982 
KOI planet candidates) from the B12 catalog included in our overall 
LC \lc \ samples. The rest of the KOI stars in the B12 catalog either fail to 
fulfill our long cadence time baseline length requirement 
(see \S \ref{sec:TFA}) or have transit depths greater than 0.04, and are 
therefore rejected by our procedure. We found that 1311 (86.4\%) of the 
KOI stars are flagged in our selection process. 1636 (82.5\%) of the KOI 
planet candidates are detected (either with the correct period reported by 
BLS or are detected with the wrong period but later recovered during 
visual inspection).
 
We failed to recover about 17.5\% of the KOI planet candidates, mostly 
due to no significant peaks found in the BLS spectra corresponding to 
the transit periods. This partly results from the choice of BLS parameters 
and also the non-periodic properties of some 
transit signals. The hard cut in DSP is the second most important reason 
for the rejection of some KOI planetary candidates. Some KOIs have low DSP 
either due to over correction of our \lcs \ or  \lcs \ with high noise level. 
There are also 40 KOI planet candidates with BLS peaks selected by our 
pipeline but rejected during visual inspection. From further investigation, 
we find that 32 of these rejected transit signals are not visible when 
folded with twice the detected periods, probably due to strong transit 
timing variations. 
A detailed detection report of all the KOIs is presented in 
Table \ref{tab:KOItable}.   
We do not model the KOIs unless the host stars are found to have other
high SNR and DSP transit signals that are not reported by the Kepler team. 
The additional signals are treated as new candidates in our short list.

\subsection{Generation of the planet candidate short list}

We examined in detail the remaining candidates which are not included 
in the publicly available lists of candidates, EBs, or false positives 
(FPs). We re-applied the cosine filter to these light curves constructed 
with a greater number of cosine functions and a smaller frequency interval. 
We aimed to clean systematic variations from the light curves (whether 
intrinsic to the stars, or due to instrumental effects), leaving 
flat \lcs\ with transits. 
The cosine functions are generated in a frequency range adjusted 
according to the visual determination of the frequency of the stellar 
variation. To preserve the transit signal, we applied a median filter 
with a window width at least twice that of the detected transit 
duration, before generating the cosine filter. BLS was 
applied again on the corrected \lcs. The first ten peaks in the BLS 
spectrum are examined and compared to the previous detection, ensuring 
the detection of the transit signal is robust. As shown below in 
\S \ref{falsepositive}, all candidates were checked against false 
positive scenarios.

%
\subsection{False positive detection and robustness checking}
\label{falsepositive}

\begin{figure}
\includegraphics[angle=0,width=0.9\linewidth]{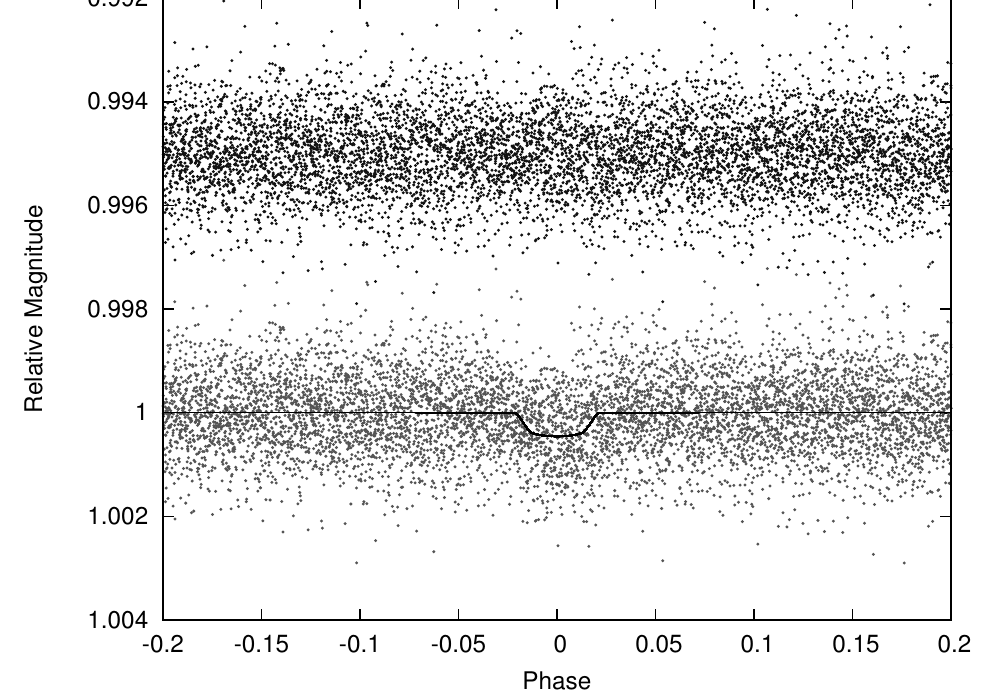} 
\includegraphics[angle=0,width=0.9\linewidth]{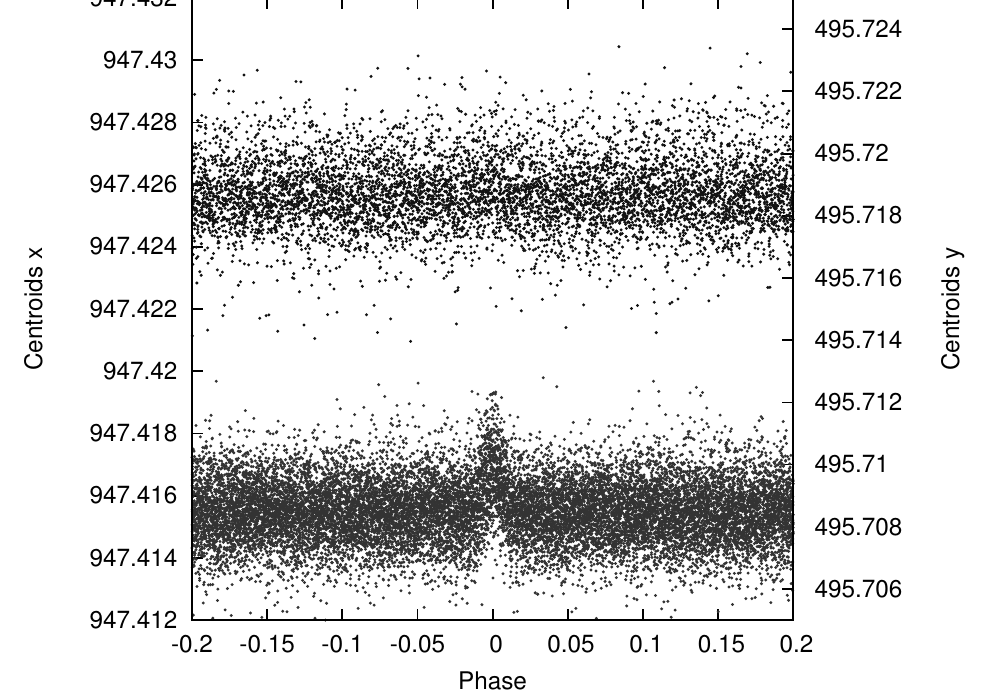} 
\caption{
Light curves and centroids for KIC~004072333, phase-folded using the
detected period and epoch. 
	Top panel: phase-folded \lc\ (bottom) with period 
$P=2.00783\,\mathrm{days}$ and $E= 2455302.8774\,\mathrm{(BJD)}$ and its 
residual from the fitting (top). 
	Bottom panel: folded centroids in the $x$ direction 
(top and grey) and the $y$ direction (bottom and dark) with the same 
parameters.
	This detection is flagged as a false positive due to the shift 
of centroids in the $y$ direction during transit.
	There is no visible companion in the 2MASS image stamp within 
$20^{\prime\prime}\times20^{\prime\prime}$ for this particular star.
\label{fig:FP}
}
\end{figure}

We use the moment-derived centroids provided by the {\em Kepler} FITS 
files to further eliminate possible FPs. For long period planet 
candidates, we visually examine the centroids at the transit time. For 
short period planet candidates, we examine the phase-folded centroid 
curves (de-trended by the cosine filter).
We present one of our ``failed'' transit candidates as an 
example in Figure \ref{fig:FP}. The folded \lc\ would naively suggest 
that there is a transit signal due to a planet with 
$R_p/\rstar=0.0198\pm0.0017$ and period of 
$P=2.00783\pm3.8\times10^{-5}$\,day. However, the phased centroids show 
a shift of $\sim0.004$\,pixels in the $y$ direction during transit 
events, which indicates this is likely to be a false positive signal 
due to a blended eclipsing binary. According to the 2MASS image stamp, 
there is no nearby source within an area of 
$20^{\prime\prime}\times20^{\prime\prime}$, i.e.~the binary is
not resolved in the 2MASS images. We provide a list of 
forty false positives flagged by this method in Table \ref{tab:fp}. 
These stars have not been reported by the {\em Kepler} team in their 
false positive lists. They are also not reported as candidates by the 
{\em Kepler} team. The stellar information and estimated shifts in 
both directions are also given in our table. 

We also used the public target pixel files and PyKE 
package\footnote{http://keplergo.arc.nasa.gov/ContributedSoftwarePyKEP.shtml}
developed by the {\em Kepler} team, to obtain pixel \lcs\ 
for all of our candidates with transit depths greater than 1\,mmag.
It is difficult to perform the same analysis on shallower transits, 
because of the low signal-to-noise of the events in the \lcs\ of 
individual pixels. 

\begin{figure*}
\includegraphics[angle=-90,width=0.9\textwidth]{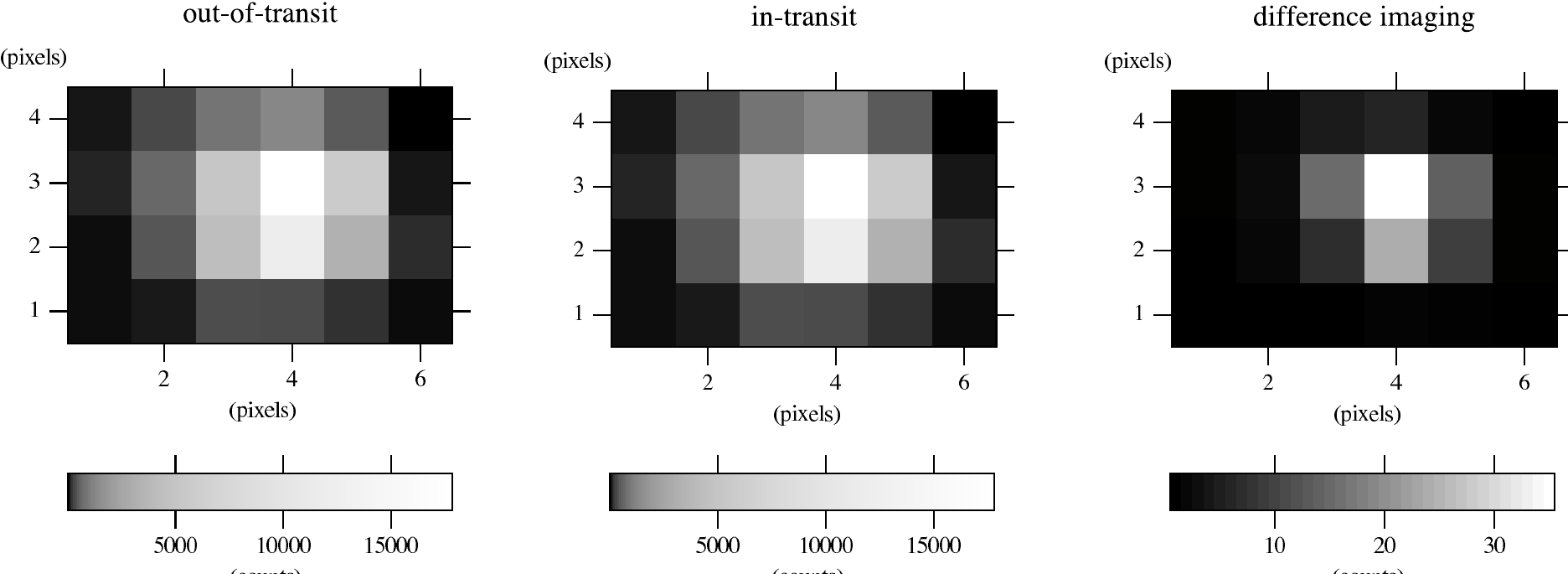} 
\caption{
Out-of-transit image (in log scale of flux), in-transit image (in log 
scale of flux) and the difference between the two (in linear scale of 
flux) from the pixel files for KIC 005437945 Q6 transit. We do not see 
a visible shift in the flux distribution on the pixel image during 
transit, demonstrating that the apparent transit is not due to a 
variation of a background source.
\label{fig:diff}
}
\end{figure*}

Here we take one of our new detections, KIC 005437945, as an example 
for our photometry analysis. We detected 4 non-periodic transit events 
altogether, which could be explained as due to two long period planet 
candidates with different epochs, depths and durations.
The transits in Q1 and Q6 are due to a $P\approx$\,440\,days
planet candidate; and the transits in Q2 and Q5 are due to a 
$P\approx$\,220\,days planet candidate (i.e.~they are in a 2:1 
resonance). We present the analysis for the transit event in Q6 here.   
We show the pixel image of out-of-transit, in-transit and the 
difference imaging in Figure \ref{fig:diff}.    
The images are computed by plotting the mean out-of-transit flux 
($\pm2\,\mathrm{day}$ around the transit events), the mean in-transit 
flux and the difference between the two. The difference imaging during 
transit is identical to the out-of-transit flux distribution. No 
obvious background source is indicated for this particular star. 

\begin{figure}
\includegraphics[angle=0,width=0.9\linewidth]{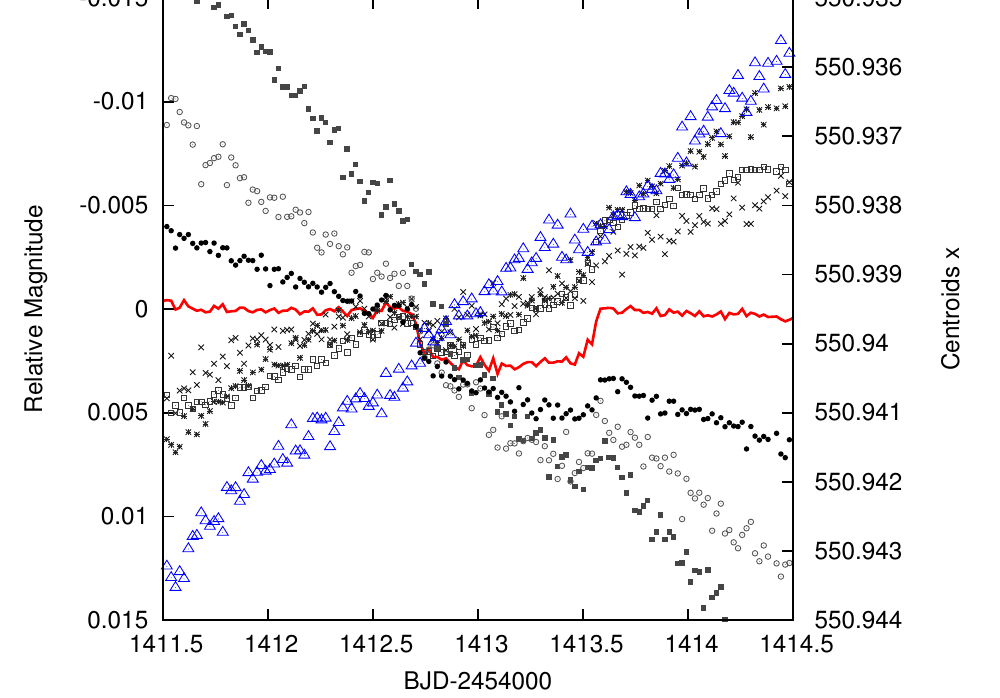} 
\includegraphics[angle=0,width=0.9\linewidth]{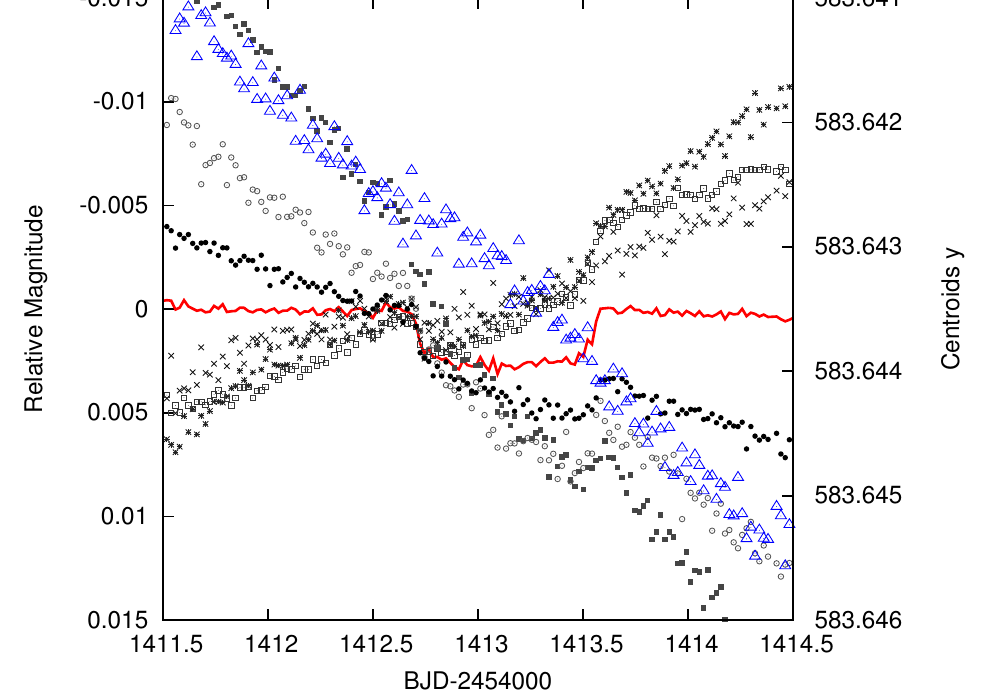} 
\caption{
	Red solid line: \lc\ flux for transit events in Q6 for 
KIC 005437945. Blue triangles: raw centroids x (top panel) and y 
(bottom panel). 
	Different types of black points: \lcs\ generated from different 
pixels separately in the aperture using kepextract from the PyKE 
package.
 \label{fig:mom}
}
\end{figure}

We further use the pixel calibration technique enabled by PyKE to 
extract \lcs\ from every single pixel in the aperture. 
Figure \ref{fig:mom} shows the \lcs\ extracted separately from 6 pixels 
in the Q6 aperture (the aperture is shown in Figure \ref{fig:diff}). We 
can see that while the magnitude in every single pixel changes during 
the transit, they all show visible evidence for the transit event with 
roughly the same depth. The averaged flux from all the pixels has a 
flat out-of-transit magnitude. In addition, the centroids do 
not present an anomalous shift during the transit events.  

For the rest of the transits which are not suitable for applying the single 
pixel analysis, we use other methods to at least ensure the robustness 
of the detection. We removed the detected transit signal according to 
the epoch, period and transit duration reported by BLS. Then we 
computed six different BLS spectra for each of the new \lcs. 
The BLS spectra were obtained with 3 different sets of parameters 
(different frequency steps and bin numbers), each set of parameters 
are used twice. The first 10 peaks of every BLS spectrum were analyzed. 
We compared these 60 frequency peaks to our original transit 
detections. For all of the planetary transit signals we analyzed, 
the original periods as well as their harmonics were not detected in the 
transit-removed \lcs.    

Altogether 150 planet candidates entered our final list of new 
detections. The properties of their host stars are provided in 
Table \ref{tab:star}. 
The detected periods range from $\sim$\,0.17\,day to $\sim$\,440\,days. 
We also found 6 new multiple systems (altogether with 15 planet 
candidates) in stars not in the B12 catalog. We found 57 new transit 
signals in KOI hosting stars; 43 of these are also independently reported 
by \citet[Hereafter O12]{Ofir:2012}. 
We also include 7 {\em single} transit events. One of these single 
transit events is around a KOI star with a known planet candidate. 
Using the convention of \citet{Batalha:2012} for single transit events, 
we assign a negative integer period number for these potential candidates. 
We compute the minimum allowed period according to the given time span 
of the \lc\ and the epoch of transit. This is taken as the estimated 
period for single transit events in modeling.

%
\subsection{Analysis}
The transit modeling of all the candidates is based only on the 
{\em Kepler} \lcs\ and the stellar parameters provided 
by the KIC. The parameters we can obtain directly from \lcs\, are the 
transit depths, durations, ingress/egress durations, and individual 
transit centers.  

Without radial velocity (RV) data, mass determinations for these 
systems are generally not available. It may be possible to measure the 
masses through subtle photometric effects, like ellipsoidal variations 
and relativistic beaming \citep{Mazeh:2010,Kipping:2011}. These effects
are prominent for close binary stellar systems as well as massive planet
companions \citep{Loeb:2003}. We did not observe these effects in our 
candidates.
 
The eccentricity is also unknown, although broad limits can be placed 
on the orbital configuration for very wide transits 
\citep{Kipping:2008}. 
An eccentric orbit could result in asymmetry in the transit \lcs, 
as well as a shift in the mid-time of transits relative to the 
occultation events.
In principle, the eccentricity could also be derived from modeling the 
detailed shape of a transit \lc. 
However, detecting these effects requires extremely high resolution and 
SNR \lcs, which were not available for our candidates. Generally 
speaking, we observe no apparent asymmetry in any of our candidates, 
which suggests modest eccentricity or certain values of argument of
periastron. We assume circular orbits in our 
modeling, following the convention of previous KOI modeling with only 
transit \lc\ information \citep{Batalha:2012}.          

We assume no flux dilution from blended nearby stars in our modeling. 
This is plausible for most of our candidates. We inspect the 2MASS image 
stamps with an area of $20^{\prime\prime}\times20^{\prime\prime}$ 
centered on each target. Four of them have nearby companions, which are 
marked out in Table \ref{tab:planet}.

In theory, one could also fit for the limb darkening coefficients, but 
this requires very high signal-to-noise and well sampled data, 
i.e.~deep transits or many transit events \citep{KippingBakos:2011}.
We use a quadratic limb darkening formalism to model the transit \lcs. 
The limb darkening coefficients (LDC) corresponding to the stellar 
atmospheres are interpolated with the stellar parameters from KIC in 
the ATLAS model grid LDCs provided by \citet{Sing:2010} for 
{\em Kepler}. The stellar parameters and the LDCs for the candidates 
are listed in Table \ref{tab:star}. We also generate a grid of 1000 
randomly selected stars from KIC with known stellar parameters and the 
interpolated their LDCs following the method described above. 
We then use these to determine LDCs for stars without information such 
as \teff, \logg\ and \feh\ by linear interpolating in the J, H, K color
space. Since the LDCs obtained with this method have larger 
uncertainties, the derived planet parameters for planets without host 
star parameters provided by KIC should be treated with caution. 

To conclude, in our transit modeling, we {\em do not} fit for:

\noindent\,a)\,the planet mass $M_p$;

\noindent\,b)\,the eccentricity $e$ of the orbit, which is assumed to 
be zero;

\noindent\,c)\,the blending due to nearby stars, which is set to be zero;

\noindent\,d)\,the LDCs of stars, which are computed as described above.

We did fit the following geometric parameters which correspond: 

\noindent\,a)\,the fractional planet radius $R_p/R_\ast$; 

\noindent\,b)\,the square of the impact parameter square $b^2$;

\noindent\,c)\,the inverse of half duration $\zeta/R_\ast=2/T_{dur}$.
This quantity is related to $a/R_\ast$ for zero eccentricity via the 
relation: 
\begin{equation}
\frac{\zeta}{R_\ast}=\frac{a}{R_\ast}\frac{2\pi}{P}\frac{1}{\sqrt{1-b^2}}.
\end{equation} 
In the fitting procedure we constrained the quantities 
$b^2$,$R_r/R_\ast$,$\zeta/\rstar$ to be the 
same for each individual transit for a selected candidate.

We also fitted for the additional parameters of out-of-transit 
magnitudes and transit centers. For the short period cases (with a period 
shorter than 30 days), we fit the median out-of-transit 
magnitude, first transit center $T_A$ and the last transit center $T_B$ 
as free parameters, with the total number of transits $N$ fixed. We 
also assume that the transits are strictly periodic. We do not model 
the out-of-transit variation; the \lc\ is assumed to be flat.
For our long period candidates, we use a slightly different method. We 
take the out-of-transit magnitude and transit centers of individual 
transits as independent parameters, the total number of free parameters 
in the fit for a transit \lc\ with $N$ transits is $2N+3$. The periods of 
the candidates with only a single transit were estimated by the stellar 
density through KIC parameters, and by assuming no limb darkening and 
circular orbits \citep{Yee:2008}. The estimated periods for these cases 
are shorter than the lower limit constrained by the fact that only one 
transit event was observed during Q1-Q6. These single events were fitted 
as if they were an individual transit in a long period system with a 
rough lower limit on the period. 

In the fitting we used the formalism of \citet{Mandel:2002}, and the 
methodology laid out in the analysis of HATNet planet discoveries 
\citep{Bakos:2010}. A 10000 step Markov Chain Monte Carlo (MCMC) 
simulation is then applied around the best fitting parameters to 
explore the parameter space. The final reported planet parameters and 
estimated errors are taken as the median and median deviation of all 
the accepted jumps in the chain. The period is then recalculated by 
taking the median of $(T_B-T_A)/N$ for all the accepted jumps in the 
chain. We derive the transit number closest to the 
average of $T_A$ and $T_B$ (weighted by their errors as derived
from the MCMC runs), and use the transit center of this event
(calculated from $T_A$, $T_B$, $N$) as the optimal epoch.

\begin{figure*} 
\centering 
\includegraphics{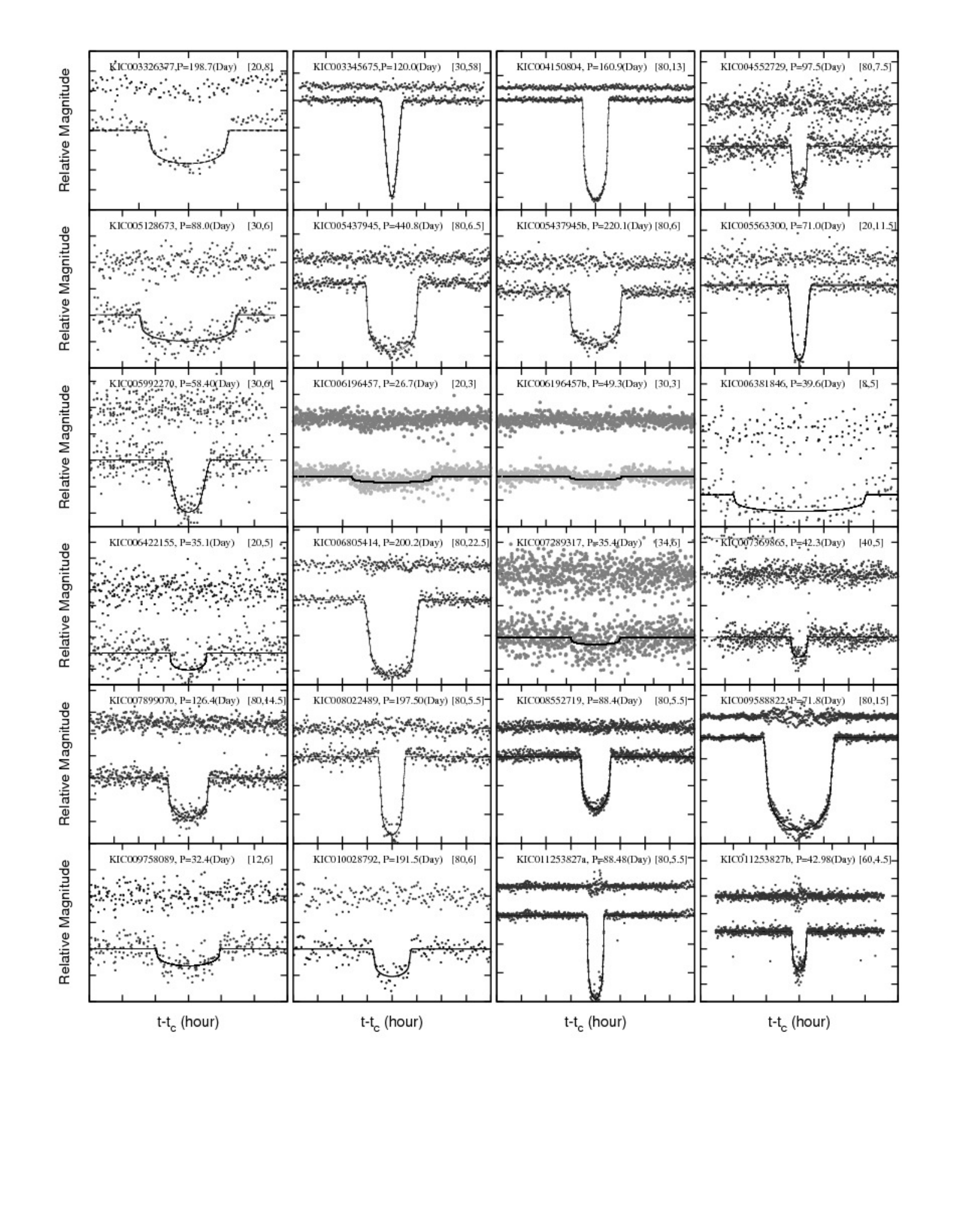}%
\caption{
New Long Period planet candidates. For each candidate we show the best-fit
model on the phase-folded \lc\ (bottom of the panel) and the residuals
(top of the panel). For each candidate, we model all transits
separately.  We then fold the transits using the locally computed
transit centers. We show the KIC number, period and scales of the 
figure of every planet candidate at the top of the subfigures. The 
x, y scales of the subfigure size are marked on the top right, 
in the units of [hours,1mmag].  
\label{fig:lptransit}
}
\end{figure*} 
\clearpage
\begin{figure*} 
\subfloat{\includegraphics[width=0.9\linewidth]{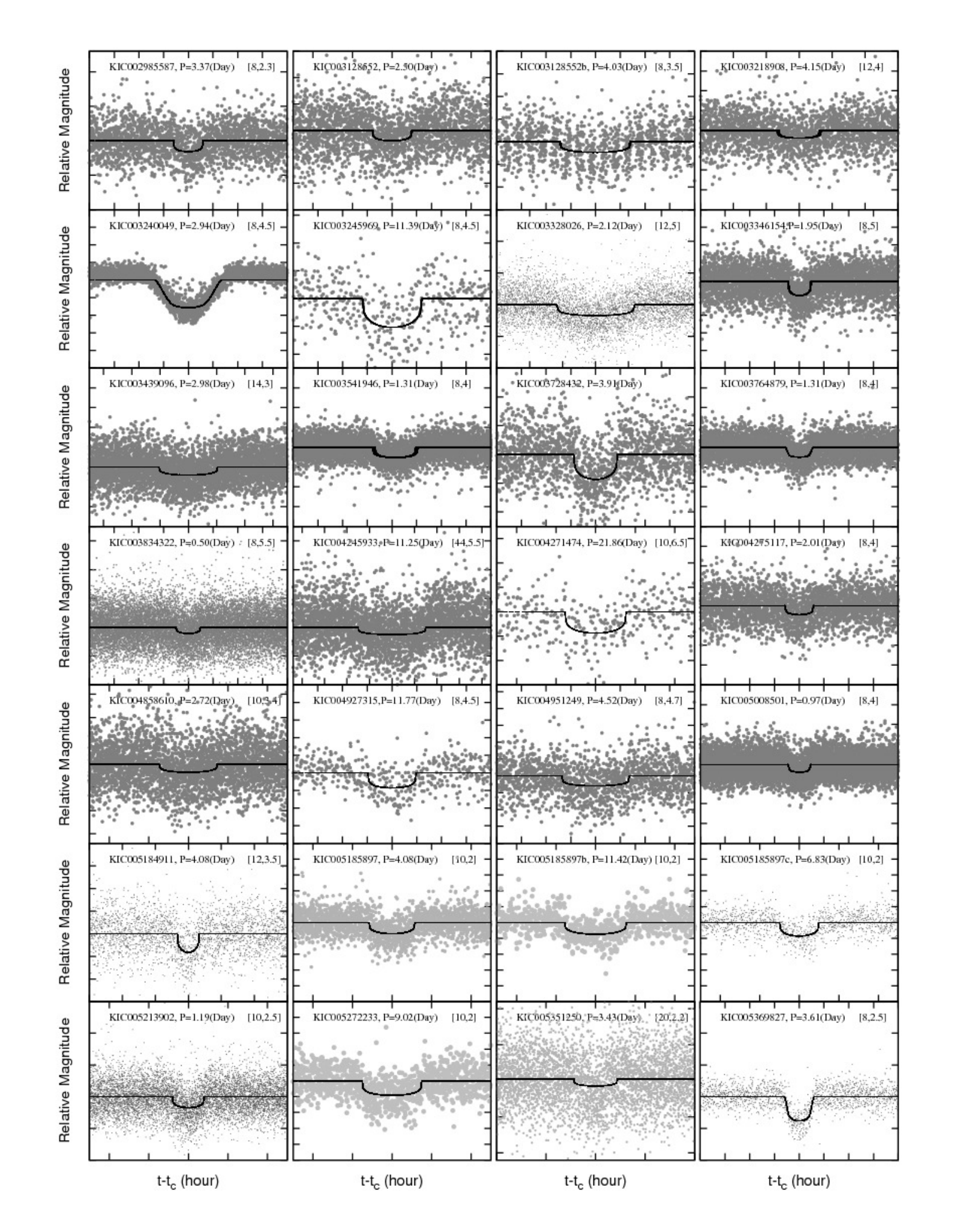}}%
\label{fig:fig8a}
\caption{(a)}
\end{figure*}
\clearpage
\begin{figure*}
\ContinuedFloat
\subfloat{\includegraphics[width=0.9\linewidth]{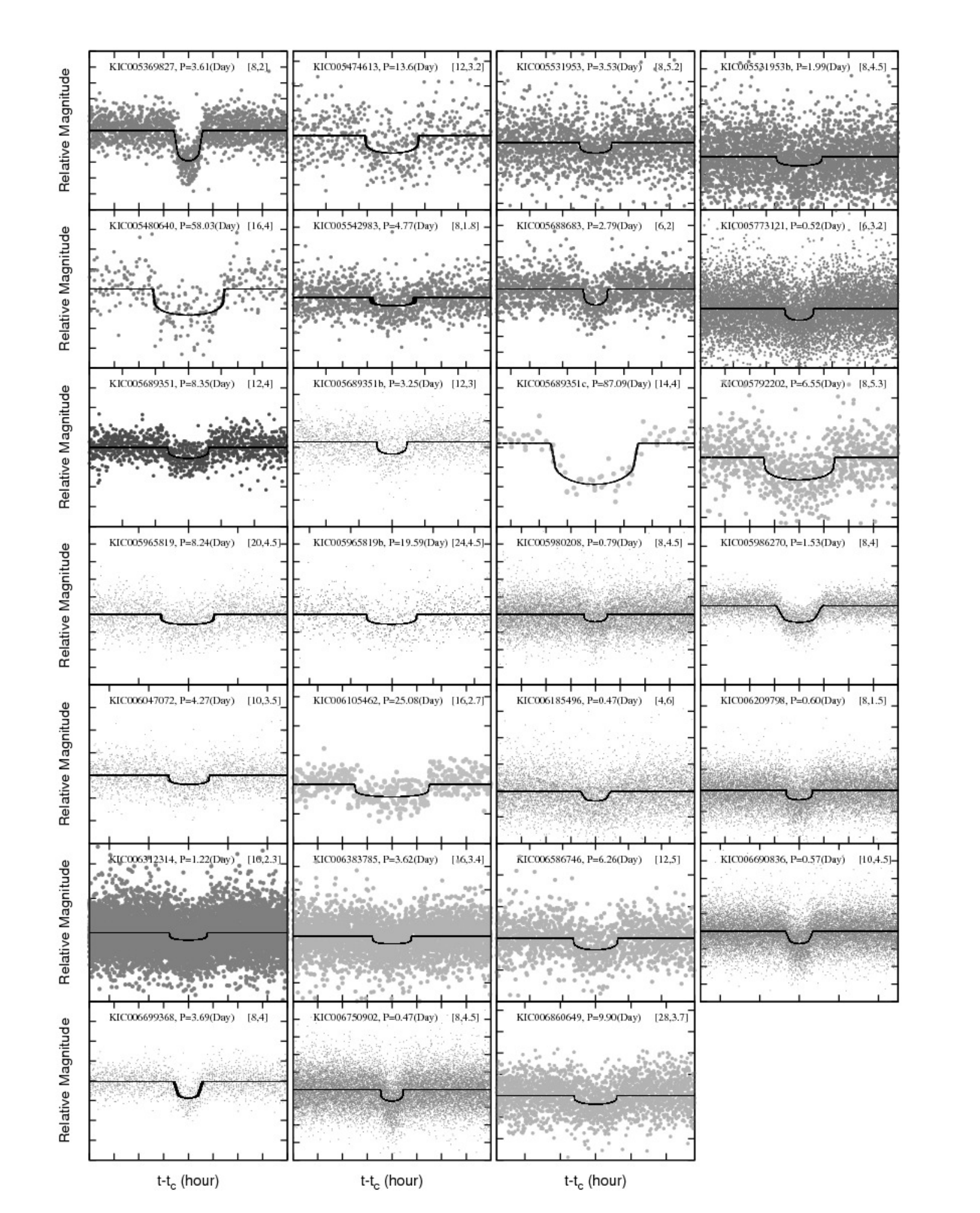}}%
\label{fig:fig8b}
\caption{(b)}
\end{figure*}
\clearpage
\begin{figure*}
\ContinuedFloat
\subfloat{\includegraphics[width=0.9\linewidth]{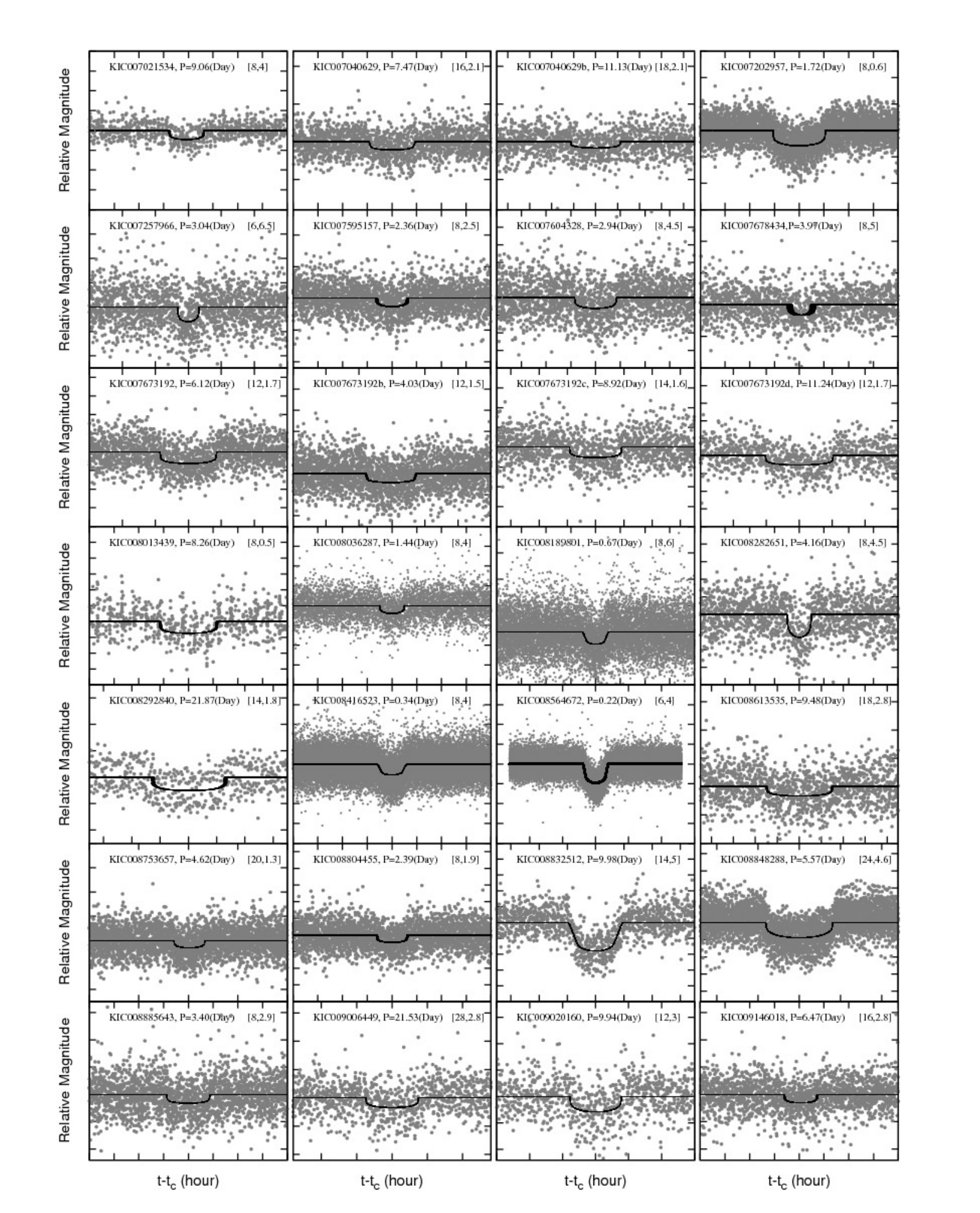}}%
\label{fig:fig8c}
\caption{(c)}
\end{figure*}
\clearpage
\begin{figure*}
\ContinuedFloat
\subfloat{\includegraphics[width=0.9\linewidth]{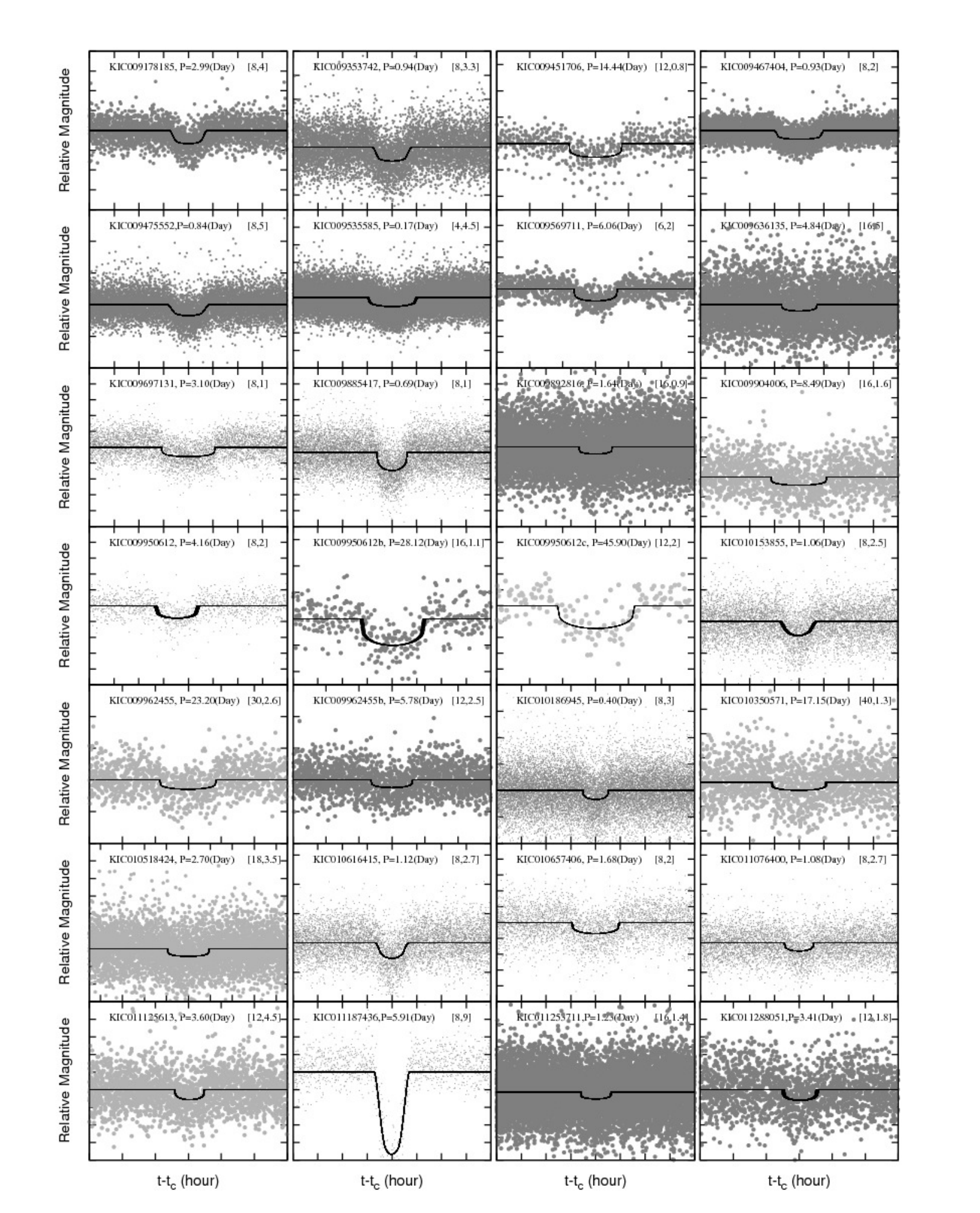}}%
\label{fig:fig8d}
\caption{(d)}
\end{figure*}
\clearpage
\begin{figure*}
\ContinuedFloat
\subfloat[][]{\includegraphics[width=0.9\linewidth]{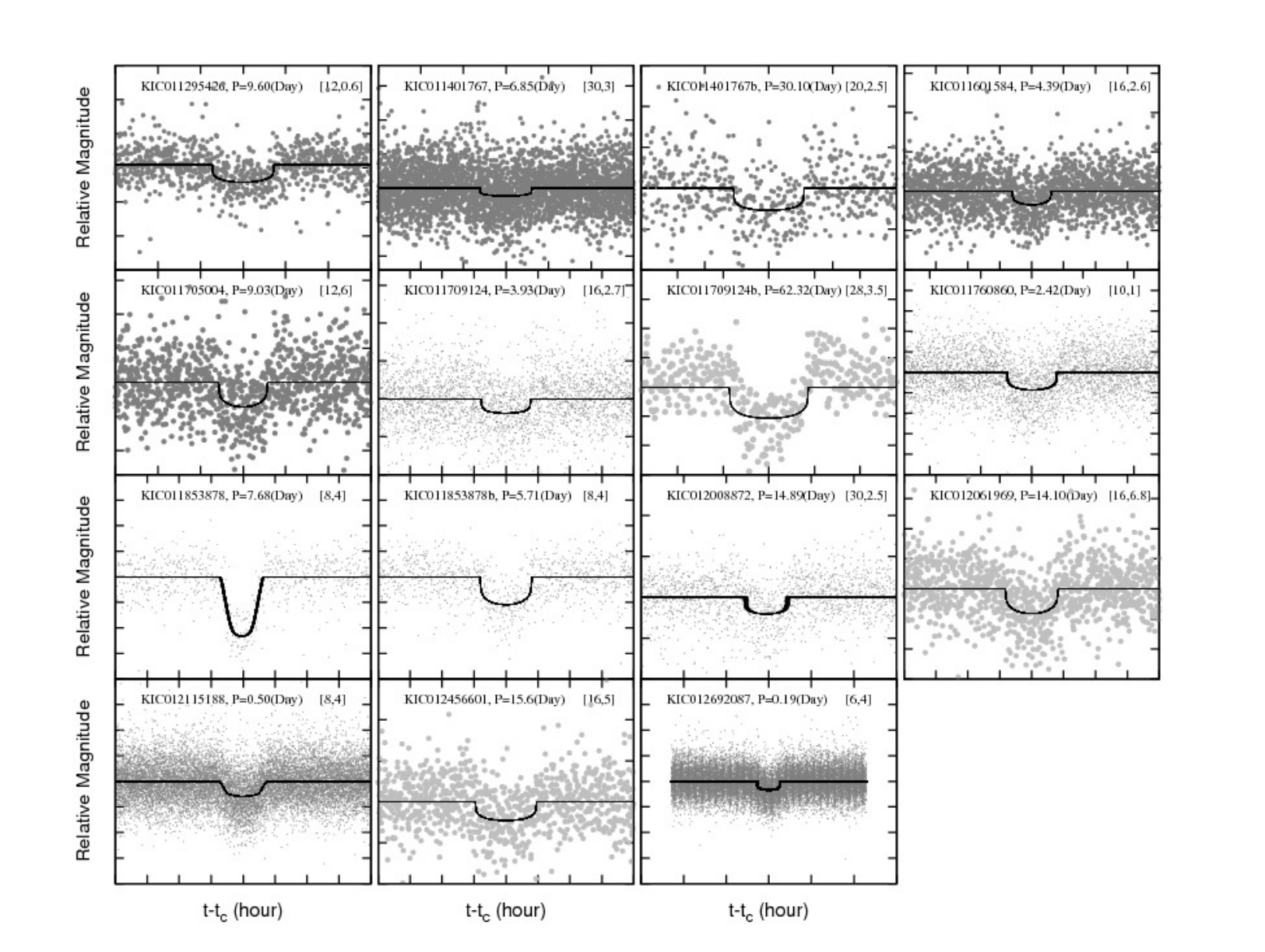}}%
\caption{
Short period planet candidates. For each candidate we show the best-fit 
model on the phase-folded \lc. The residuals are flat so we don't show 
them in the figures. The x, y scales of the subfigure size are marked 
on the top right, in the units of [hours,1mmag].  
\label{fig:sptransit}
}
\end{figure*}

\begin{figure*} 
\centering 
\includegraphics{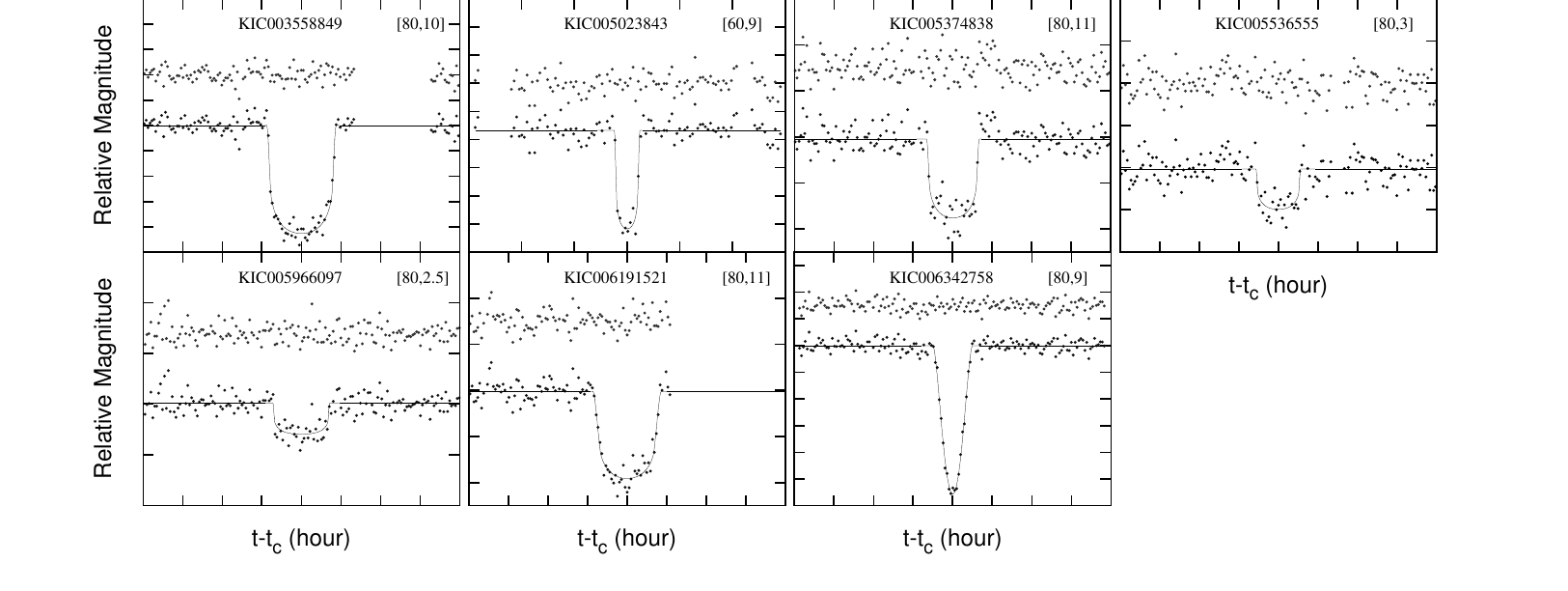}

\caption{
Single transit events. For each candidate we show the best-fit model 
on the \lc, together with the residuals from the fit (the configuration 
is the same as in Figure~\ref{fig:lptransit}). The x, y scales of the 
subfigure size are marked on the top right, in the units of [hours,1mmag].  
\label{fig:singletransit}
}
\end{figure*}
\clearpage
\clearpage

\subsection{Properties of candidates}

We summarize the planet parameters for all the candidates in 
Table \ref{tab:planet}. The best fitted model on the phase folded \lcs\ 
and the residuals from the best fitted model are presented in Figures 
\ref{fig:lptransit}, \ref{fig:sptransit} and \ref{fig:singletransit} 
for long period, short period and single transit planet candidates.

\begin{figure*}
\includegraphics[angle=0,width=0.45\linewidth]{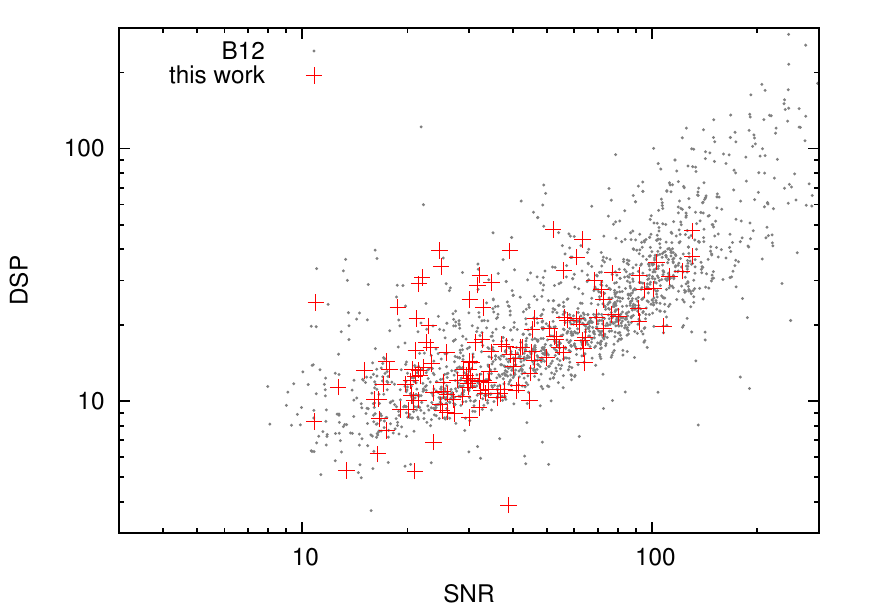} 
\includegraphics[angle=0,width=0.45\linewidth]{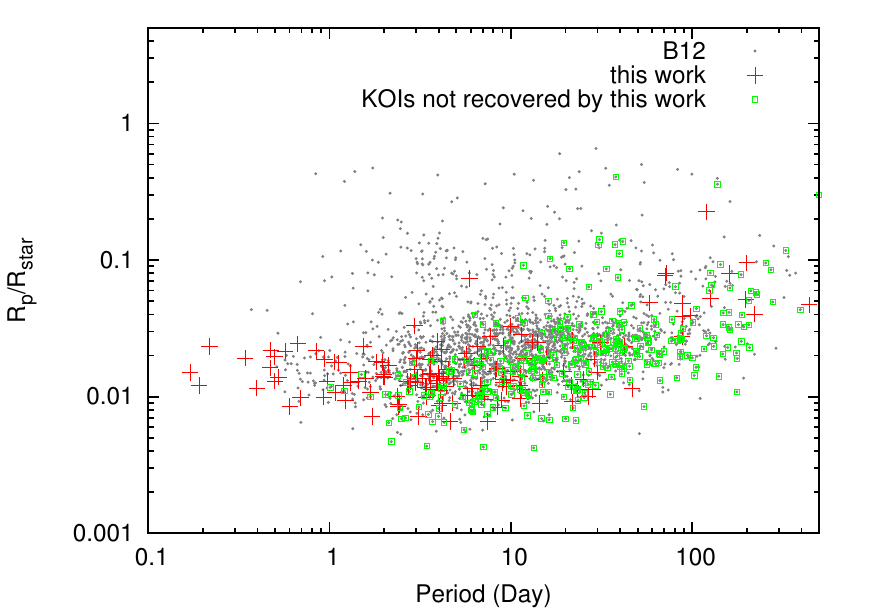} 
\caption{
Comparison of KOIs and the planet candidates reported in this paper.
Left: in SNR-DSP space; Right: in period-transit depth space. The 
black dots represents all the KOI planet candidates. The red cross 
represents the new candidates in this paper. In the right figure,
the green squares show the KOIs missed by our searching process. 
\label{fig:compare}
}
\end{figure*}

We compare the planet candidates reported in this paper with the B12
catalog KOIs in Figure \ref{fig:compare}. Our new planet
candidates follow a similar distribution in the SNR and DSP space as the
KOIs. This demonstrates that our recovery of new candidates is due to 
the efficiency of our pipeline rather than due to a reduction in the 
selection thresholds. We also show in the right of Figure \ref{fig:compare} 
that the majority of new candidates have short period (less than 10 days) 
and small transit depth. For comparison, we plot the KOIs missed by 
our search in green, suggesting that our detection sensitivity drops as 
the period of the transit signal increases.      
 
\begin{figure*}
\includegraphics[angle=0,width=0.45\linewidth]{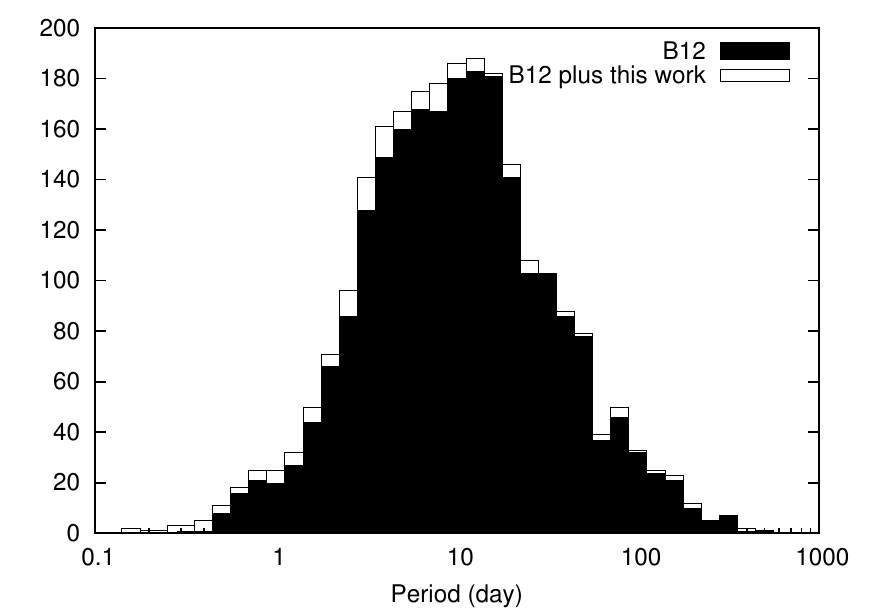} 
\includegraphics[angle=0,width=0.45\linewidth]{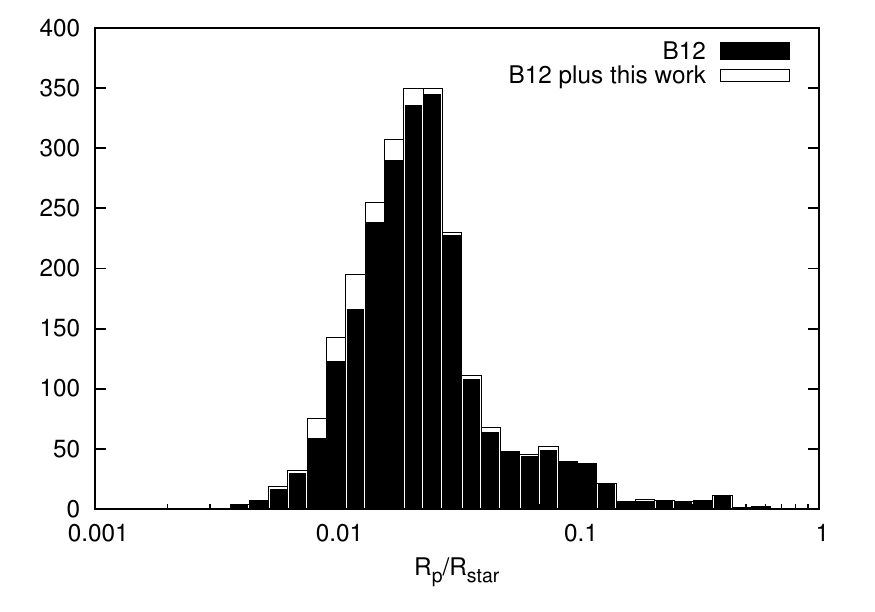} 
\caption{
Comparison of the distribution of periods (left) and $R_p/R_{star}$ 
(right) between KOIs and the total (KOIs and this work). The shadowed
regions is the distribution from KOI planet candidates alone.
\label{fig:hst}
}
\end{figure*}

We plot the period and planet radius (in units of host star radii) 
histogram in Figure \ref{fig:hst}. The turn over points in both 
the period and radius space are modified with the supplementation of
our new candidates. We suggest that our methods are more sensitive to
short period candidates since the detection efficiency of BLS is generally
higher in high frequency. We also confirm that the KOI samples are 
almost complete in the mid-period range. We do not have much advantage in 
the long period range over the original Kepler pipeline. The sample size 
around $P\sim100$ days slightly gains with our new candidates. Signals with 
periods longer than 100 days usually suffer from over-correction of 
individual transit or low DSP simply because of lack of points in the 
transit phase.    


Finally, we provide remarks on some of the interesting systems  
below \footnote{The planet and stellar parameters we quote here are only
approximated numbers for easy comparison. For accurate parameters and
estimated errors, refer to Table \ref{tab:planet}.}.

{\bf KIC\,005185897:} We found three planet candidates in the system, 
all with transit depths $\sim 0.1 {\rm mmag}$. The periods of 
the candidates are 4.08 days (a), 11.42 days (b) and 6.83 days (c).
(See Figure \ref{fig:sptransit} (a), row 6, column 2, 3 and 4.)   
The stellar radius from KIC is only 0.51 \rsun, which makes the the 
modelled radii extremely small for all the candidates in the 
system. The $P=11.42$ day signal and 4.08 day signal are modelled to be 
$\sim$0.62 \rearth, and the 6.83 days signal is slightly larger 
(0.67\rearth). They are also the smallest planet candidates we found 
around non-KOI stars. Both the pair b and c, c and a are around 5:3 
resonance. The planet candidate (a) was also found by the TPS 
algorithm as a potential transit signal (T12).

\begin{figure}
\includegraphics[angle=0,width=0.8\linewidth]{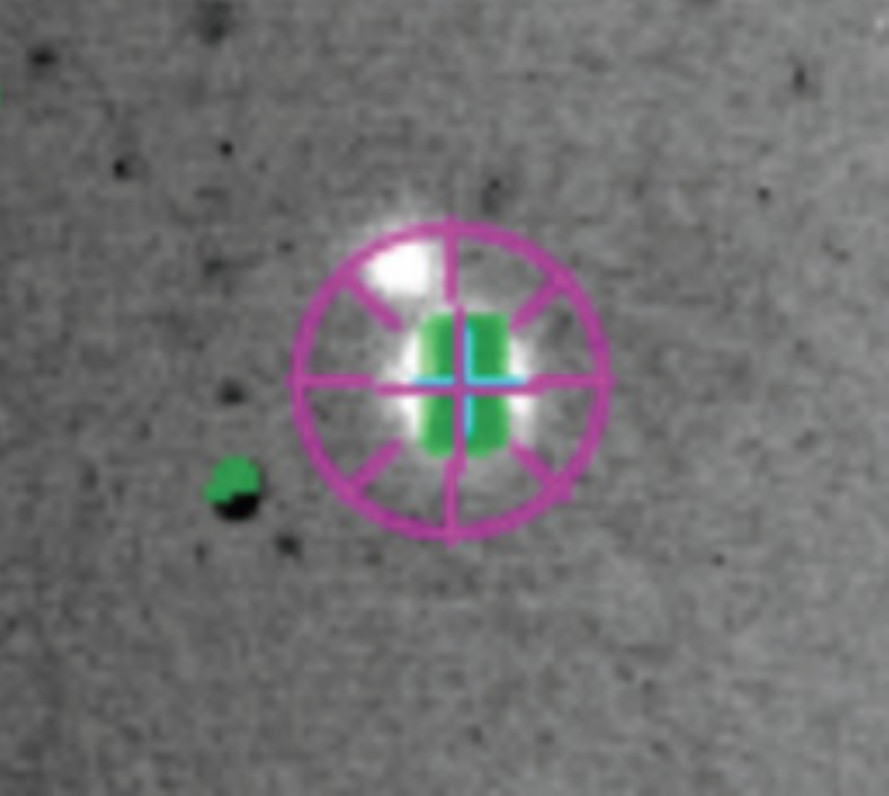}
\caption{
	Image of the star KIC 5437945 taken with the APO 3.5\,m Echelle 
slitviewer. The guider is centered on the star. A companion is resolved 
within $\sim\,5^{\prime\prime}$.  
\label{fig:K54photo}
}
\end{figure}
\begin{figure}
\includegraphics[angle=0,width=\linewidth]{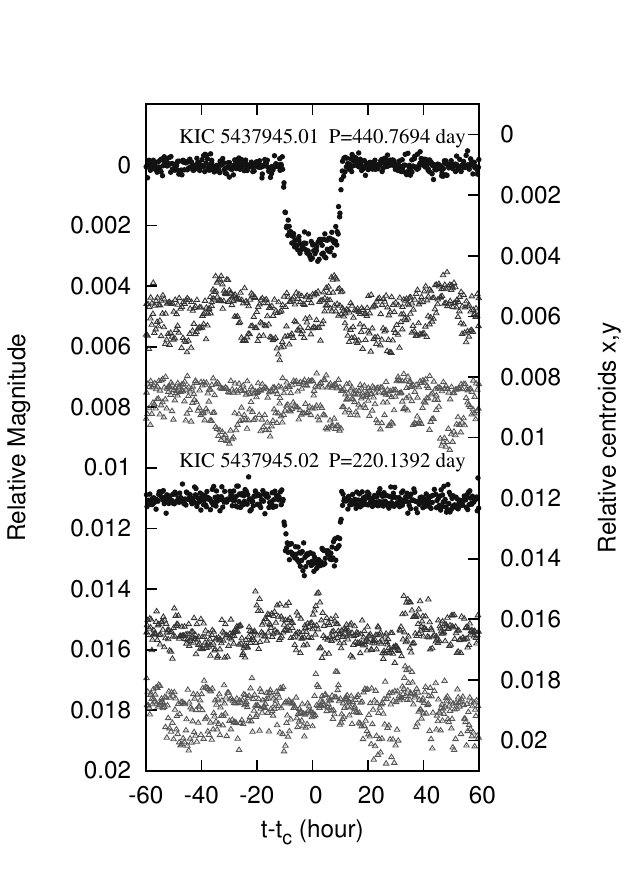}
\caption{
	The two planet candidates of star KIC 5437954: from black to 
grey are the phased folded \lc, and the phased folded x and y 
direction centroids, all displayed separately. 
The upper (lower) three plots are for the first (second) planet 
candidate, with period $P\sim$ 440(220)\,day. The centroids in the 
figure are the mean centroids after de-trending.   
\label{fig:K54mom}
}
\end{figure}
\begin{figure}
\includegraphics[angle=0,width=0.9\linewidth]{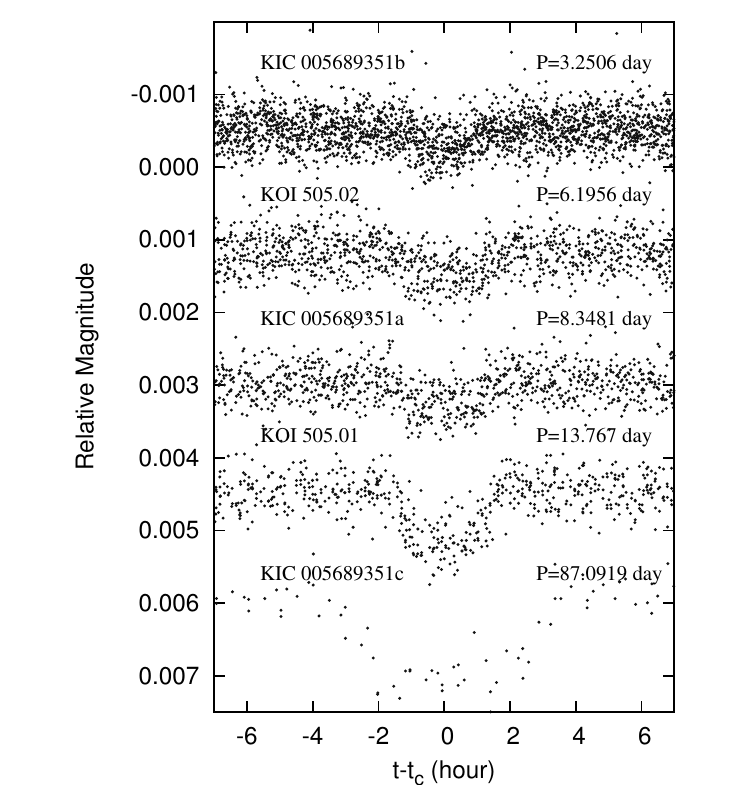} 
\caption{ 
	Folded de-trended \lc\ of KIC 005689351. Each transit is folded 
relative to its measured epoch, with the signal corresponding to the 
other planet candidates removed from the \lc. From top to bottom, the 
periods of the planet candidates are 3.25 days, 6.20 days, 8.35 days, 
13.77 days and 87.09 days, respectively. The candidates with periods of 
13.77\,days and 6.20\,days have already been identified as KOI 505.01 
and KOI 505.02.
\label{fig:multi}
}
\end{figure}

{\bf KIC\,005437945:} We identify four transit events in the \lc. 
The transit events in Q1 and Q6, and the ones in Q2 and Q5 share the 
same depths and durations, respectively. The former is modelled to be a 
7.5\,\rearth\, planet candidates with 440\,day period; the latter is 
modelled to be a 6.4\,\rearth\, planet candidate with a period 
of 220\,days. (See Figure \ref{fig:lptransit}, row 2, column 2 and 3.) 
We note that the transit in Q1 is independently identified and 
classified as a single transit event by the Planet Hunters 
\citep{Schwamb:2012}. A tweak happened during the transit in Q2 for the 
inner planet candidate; by carefully offsetting the magnitude on both 
sides of the tweak, we recover the full transit signal. 
We resolved a faint close companion ($\sim5^{\prime\prime}$) 
in the Echelle slitviewer of the APO 3.5\,m telescope (See Figure 
\ref{fig:K54photo}). We examined 2MASS image stamps in J,H and K and DSS 
1,2 images in red and blue. All of the above show that the companion is 
bluer than KIC\,005437945, which indicates that it is unlikely to be a 
physically associated companion. We present the phase folded de-trended 
centroid for both planet candidates in Figure \ref{fig:K54mom}.
No anomalous motion is shown in either direction during transit.         

{\bf KIC\,005965819:} We found two planet candidates in the system, 
both with modelled radii less than 2\,\rearth. The inner planet 
candidate orbits the host star with a period of 8.2\,days; the outer 
planet candidate has a 19.6\,day period. (See Figure \ref{fig:sptransit} (b), 
row 3, column 1 and 2.) They are not in a tight resonance with each 
other. We note that the transit durations of these two planet 
candidates are longer than the values expected assuming circular 
orbits. This might indicate that the planets are in eccentric orbits 
or the stellar radius of KIC\,005965819 is in fact larger than what is 
listed in the KIC.   

{\bf KIC\,007673192:} We found four planet candidates in the system, 
all with modelled radii around 1\,\rearth. The periods of the 
candidates are 6.12 days (a), 4.03 days (b), 8.92 days (c) and 
11.24 days (d). (See Figure \ref{fig:sptransit} (c), row 3.) They are not 
around any low order resonance pairs. T12 also identified the P=6.12 day 
signal as a potential transit signal.   

{\bf KIC\,009962455:} We found two planet candidates in the system, 
both with modelled radii around 1\,\rearth. The periods of the 
candidates are 23.20 days (a), 5.78 days (b), in 4:1 resonance. 
(See Figure \ref{fig:sptransit} (d), row 5, column 2 and 3.)
The 23.20 day component in the system was also found by T12. 

{\bf KIC\,009535585:} The shortest period candidate we found. The 
signal is modelled as a 1.6\,\rearth\ super earth with a period of 
only 0.17 day. (See Figure \ref{fig:sptransit}(d), row 2, column 2.) If 
this candidate is confirmed, it would hold the record of the shortest 
period among the KOIs.   

{\bf KIC\,011253827:} We found two planet candidates in the system. 
The outer planet is Saturn-like with a radius of 6.18 \rearth\ and period 
of $\sim$ 88 days. The inner planet is $\sim$ 3.82 \rearth\ with a 43 
days period. (See Figure \ref{fig:lptransit}, row 6, column 3 and 4.) 
The Planet Hunters independently found these signals in their web 
discussion.   

{\bf KIC\,012692087:} Also a extremely short period candidate, really 
similar to KIC\,009535585, with 0.19 day period and 1.5 \rearth\ radii. 
(See Figure \ref{fig:sptransit}(e), row 4, column 3.)  

{\bf KIC\,005689351 (KOI\,505):} 
A rich system with potentially as many as 3 more planet candidates. 
We initially identified a 2.2\,\rearth\ planet candidate with a period of 
8.348 (a) day in the system, which already has 2 KOIs. While this paper was 
under revision, O12 pointed out that there might be 2 additional 
signals in the system, with periods of 3.25 days (b) and 87.09 (c) days. 
(See Figure \ref{fig:sptransit} (b), row 3, column 1, 2, 3.) We find the same 
signals in this \lc\ after reprocessing the data. The folded \lc\ of 
the three planet candidates are separately shown in Figure \ref{fig:multi}. 
When modeling one planet candidate, we make use of the detected periods 
and epochs to filter out the transit signals due to the other two. The 
top and middle panels present the two KOIs, with periods of 13.767 days 
and 6.1956 days. The third planet candidate we found is in a 4:3 
resonance with the latter. This is the first reported 3:4:6 resonance 
3-planet candidate system. \citet{Rein:2012} suggests that a resonance 
chain of three or more planets, in analog to Jupiter's moon system, 
would overcome the difficulties of forming the 4:3 resonance in the 
traditional way. This system provides an interesting testing-ground for 
the theory.

{\bf KIC\,007595157 (KOI\,246):} A sub-earth candidate with 
0.64 \rearth\, orbiting the star with a period of $\sim$ 2.4 days. 
(See Figure \ref{fig:sptransit} (c), row 2, column 2.)
It was a single planetary system in B12 catalog, with KOI\,246.01 
(1.3 \rearth\ radii and 3.4 days period). 

{\bf KIC\,008753657 (KOI\,321):} A candidate hot Mars, also the 
smallest candidate from our new findings, with a modelled radius of 
0.5 \rearth \ and period of 4.6 days. (See Figure \ref{fig:sptransit} (c), 
row 6, column 1.) This is also reported by the Ofir 
team (O12) with similar modelled parameters. The system was previously 
known to host KOI\,321.01, a 2.4~days candidate superearth. 

{\bf KIC\,11295426 (KOI\,246):} B12 catalog found it to host 
a superearth (KOI\,246.01) with a 5.38 days period. We found an outer 
planet candidate with 0.58 \rearth \ radii and 9.6 day period. 
(See Figure \ref{fig:sptransit} (e), row 1, column 1.)
This is also reported by O12.

\section{Conclusions}
\label{sec:Conc}
We have analyzed 124840 stars with public {\em Kepler} data from
quarters Q1--Q6 in total. The large majority of our candidates have 
already been identified by the {\em Kepler} team. 
We recover 92\% of Kepler findings from the B11 catalog and the 
majority (86.4\% of the planet hosting stars and 82.5\% of the planet 
candidates) of the B12 catalog. Forty new false positives are identified 
in our analysis of centroids variation. We report 150 new planet 
candidates and 7 {\em single} transit events that haven't been 
assigned as KOIs (Kepler Objects of Interests) in this blind search. 
55 of these planet candidates are listed as potential transit signals 
by the automatic TPS algorithm developed by the {\em Kepler} 
team (T12). While this paper was under revision, O12 conducted a search of 
new planet candidates using the SARS pipeline in all the KOIs, 43 of 
the planet candidates in this work overlap with their findings. To our 
best knowledge, 22 of these planet candidates and 3 of the single 
transits are also independently identified by the Planet Hunters in 
their public website discussion session. 40 of the planet candidates 
and 4 of the single transits are reported for the first time.

The periods of our new candidates range from $\sim$\,0.17\,days 
to $\gtrsim$\,440\,days. The estimated planetary radii vary from 
$\sim$0.5\,\rearth\ to 62\,\rearth. 124 of the planet candidates and 
three of the single transits have sizes smaller than 3\,\rearth. We also 
found 6 new candidate multiple systems. In addition, we found 57 more 
planet candidates in the already known KOI systems. By comparing our new 
findings with the Kepler candidates, it appears that for short periods 
($<10$ day), our pipeline and search procedure finds somewhat more 
candidates than the Kepler pipeline. Most of the {\em single} transit 
events (and a few of the planet candidates) were found by visual inspection of 
the \lcs\ that were flagged by BLS. This suggests that combining 
automated searches, and visual inspection is an efficient approach 
for transiting planet searches. Our searching process could potentially 
act as a supplementation of the findings from the Kepler team to improve 
the statistics in the short period planets. We trust that independent 
searches will be benefit both the Kepler team and the community.

\clearpage

\begin{deluxetable}{cccc} 
\centering
\tablewidth{0pc}
\tablecaption{Anomaly Summary Table \label{tab:anomaly}} 
\tablehead{ 
\colhead{Start(BJD-2454000)} & \colhead{End(BJD-2454000)} & 
\colhead{Quarter} & \colhead{Anomaly Type}
} 
\startdata
\input{anomaly.tex}
\enddata
\tablenotetext{a}{
	EXCLUDE: manually excluded cadence before pipeline processing.
} 
\tablenotetext{b}{
	SAFE MODE: due to unanticipated sensitivity to cosmic 
radiation, or unanticipated responses to command sequences. The 
amplitude of flux is affected after the end date of safe mode, data for 
additional 3 days are removed rather than corrected. Data on both side 
are offset in magnitude to ensure continuity by a polynomial fit.
}

\tablenotetext{c}{
	TWEAK (Attitude Tweaks): discontinuities in the data due to
small attitude adjustments. The discontinuities are corrected by
offseting the data on both sides using a polynomial fit.
}

\tablenotetext{d}{
	COARSE POINT (Loss of Fine Pointing): due to losing fine 
pointing control, removed.
}

\tablenotetext{e}{
	EARTH POINT: change of attitude due to monthly data downlink.
Affect data the same way as safe mode. Corrected by the same method as 
safe mode.
}

\tablenotetext{f}{
	ARGABRIGHTENING: diffuse illumination of the focal plane. 
Removed. Only argabrightening events longer than one cadence are listed 
here.
 }

\tablenotetext{g}{
	MEMORY ERROR: due to onboard spacecraft errors, gapped by the
        pipeline, the continuity on both ends of a gap is
        checked. Only a single memory error longer than one cadence is
        listed here.
}

\tablenotetext{h}{
	The format of data release handbook 8 for Q5 data is different 
from other quarters. There are no safe mode, tweak, coarse point or 
exclude phenomena in Q5, only two big gaps, possibly due to earth 
point, are listed.
}
\end{deluxetable}

\clearpage 
\begin{landscape}
\begin{deluxetable}{lcccrrrrrrc}
\centering
\tablewidth{0pc}
\tablecaption{Detection report of KOIs\label{tab:KOItable}\tablenotemark{h}}
\tablehead{\colhead{KOI} & \colhead{KIC} & \colhead{Kepperiod} & \colhead{period} & \colhead{Kepepoch} & \colhead{epoch} & \colhead{dip} & \colhead{Q\tablenotemark{i}} & \colhead{SNR} & \colhead{DSP} & \colhead{COMMENTS} \\
\colhead{} & \colhead{} & \colhead{day} & \colhead{day} & \colhead{BJD-2454000} & \colhead{BJD-2454000} & \colhead{magnitude} & \colhead{} & \colhead{} & \colhead{} & \colhead{}}
\startdata
\input{KOI_short.tex}
\enddata
\tablenotetext{a}{
sod (short of data): KOIs don't have sufficient length of long cadence data
to fulfill the requirement of TFA. They are not selected for our reanalysis.
}
\tablenotetext{b}{
wp (wrong period): KOIs recovered with a wrong period. For some planet 
candidates, the harmonics of the KOI period (or other periods due to the 
imperfect correction of the light curve) show higher SNR and DSP in the 
BLS analysis. The SNR and DSP value reported here corresponds to the detected 
signal instead of the KOI period. In other cases, frequencies not 
related to the signal are detected, the true signal is recovered in 
visual inspection, we fill the SNR and DSP with '-'.  
}
\tablenotetext{c}{
snd (single transit, no detection): KOIs with a single transit, and all the 
reported BLS peaks are under our selection limits. 
}
\tablenotetext{d}{
ld (low DSP): KOIs with lower DSP than our selection limits, therefore not 
recovered in our selection process. 
}
\tablenotetext{e}{
lr (large radius): KOIs with transit dip larger than 0.04, therefore not 
selected for analysis.
}
\tablenotetext{f}{
ls (low SNR): KOIs with lower SNR than our selection limits, therefore not 
recovered in our selection process. 
}
\tablenotetext{g}{
nd (no detection): KOIs for which the first five BLS peaks are under
our selection limits, and the correct period is not one of the first
five BLS peaks.
}
\tablenotetext{h}
{
We compare the parameters of KOIs with the B12 catalog. The period, 
epoch, dip and q are taken from our BLS analysis instead of modeling the 
candidates. A complete version of this table can be accessed from the 
online version and download from https://sites.google.com/site/largepunkelephant/tables.  
}
\tablenotetext{i}
{
Q: a dimensionless parameter representing the transit duration. 
Defined as $T_{\rm dur}/P$. 
}
\end{deluxetable}
\end{landscape}
\clearpage
\begin{deluxetable}{lcccrrrrr}
\centering
\tablewidth{0pc}
\tablecaption{False Positive List\label{tab:fp}}
\tablehead{
\colhead{Star}  & \colhead{Kepmag} & \colhead{RA} & \colhead{Dec} 
& \colhead{Epoch} & \colhead{Period} & \colhead{Displacement x\tbn{b}} 
& \colhead{Displacement y\tbn{b}} & \colhead{SNR} \\
\colhead{} & \colhead{} & \colhead{(deg)} & \colhead{(deg)} 
& \colhead{(BJD-2454000)} & \colhead{(day)} & \colhead{(Pixel)} 
& \colhead{(Pixel)} & \colhead{}
}
\startdata
\input{fp_short.tex}
\enddata 
\tablenotetext{a}{
	We only report the false positives not in the {\em Kepler} FP 
catalog here. We do not report FPs in {\em single} transit events. 
These FPs are all identified by examining the centroids. We 
refer to the text for a detailed description of the method. A complete 
version of this table can be accessed from the online version and 
downloaded from https://sites.google.com/site/largepunkelephant/tables.  
}
\tablenotetext{b}{
	The flux centroid displacement amplitude is computed from the
        phase folded centroids. The x and y direction are listed
        separately. We do not list a displacement when the shift is
        not detected.
}
\tablenotetext{c}{
	There are visible companion(s) in the 2MASS image stamp within 
$20^{\prime\prime}\times20^{\prime\prime}$.
}
\end{deluxetable}
\clearpage

\begin{deluxetable}{cccccccc}
\tablewidth{0pc}
\tablecaption{Stellar Parameter Table 
\label{tab:star}}
\tablehead{
\colhead{Star}  & \colhead{Kepmag} & \colhead{\teff} 
&\colhead{\logg} &\colhead{\feh} & \colhead{$R_{\star}$} & 
\colhead{$u_a$\tablenotemark{b}} & \colhead{$u_b$\tablenotemark{c}} \\
\colhead{} & \colhead{} & \colhead{(\rsun)} & \colhead{(K)} 
& \colhead{(cgs)} & \colhead{} & \colhead{} & \colhead{}
} 
\startdata
\input{stellar_short.tex}
\enddata
\tablenotetext{a}{
A complete version of this table can be accessed from the online version of the paper 
and downloaded from https://sites.google.com/site/largepunkelephant/tables.
} 
\tablenotetext{b}{
	The quadratic limb darkening parameter a. 
}
\tablenotetext{c}{
	The quadratic limb darkening parameter b.
}
\tablenotetext{d}{
	The limb darkening coefficient is obtained from the estimated
        stellar properties based on the J, H and K magnitudes.
}

\end{deluxetable} 
\clearpage

\begin{landscape}
\begin{deluxetable}{lrrrrrrrrrcrcc}
\tabletypesize{\scriptsize}
\centering
\tablewidth{0pc}
\tablecaption{Planet Candidates Table\label{tab:planet}}
\tablehead{
\colhead{KIC}  & \colhead{Epoch} &\colhead{$\sigma_E$} &\colhead{Period}
 & \colhead{$\sigma_P$} & \colhead{$R_p$} & \colhead{$R_p/R_\ast$} 
& \colhead{$\sigma_R$}&\colhead{$b$} & \colhead{$\zeta/R_\ast$} 
& \colhead{SNR} & \colhead{DSP} & \colhead{$\chi^2$}  & \colhead{Comments}\\ 
\colhead{}  & \colhead{(BJD-2454000)} &\colhead{($10^{-3}$)}
&\colhead{(day)} & \colhead{($10^{-4}$)}&\colhead{(\rearth)} 
& \colhead{} & \colhead{$(10^{-3})$}&\colhead{} & \colhead{($day^{-1}$)} 
& \colhead{} & \colhead{} & \colhead{} & \colhead{}
} 
\startdata
\input{planet_short.tex}
\enddata 
\tablenotetext{a}{
 A complete version of this table can be accessed from the 
online version and downloaded from https://sites.google.com/site/largepunkelephant/tables.
}
\tablenotetext{b}{
  Stars already identified as KOIs. The candidates presented here are
  transit signals that have not previously been detected in these
  systems.
}
\tablenotetext{c}{
  Planet candidates also identified by O12. 
}
\tablenotetext{d}{
 Planet candidates classified as potential transit candidates by T12.
 If noted as T12t, the same signal is identified as a different set 
of parameters by T12.
}
\tablenotetext{e}{
 The signal is selected by a different period originally and then 
corrected by visual inspection. The SNR and DSP reported here 
correspond to the period we report here instead of the 
selected period. 
}
\tablenotetext{f}{
Planet candidates classified as potential transit candidates by 
PlanetHunters. If noted as PHt, the transit feature is identified by 
PlanetHunters but the they did not report a period for comparison.
These information can be accessed through http://www.planethunters.org. 
}
\tablenotetext{g}{
  Systems may be blended by nearby stars. There exists a visible 
companion(s) within 20 square arcseconds in a 2MASS image stamp 
centered on the target star.   	
}
\end{deluxetable} 
\end{landscape}
\clearpage

\section*{Acknowledgments}

We thank the {\em Kepler} team for the high quality public data, as well as
the PlanetHunter team for their helpful comments.  We also thank the
anonymous referee for the insightful comments that greatly improved this
paper.  This research was partly funded by NSF grant AST-1108686.

%

%
\end{document}

%% file: newcommand_gen.tex
\newcommand{\forloop}[5][1]%
{%
\setcounter{#2}{#3}%
\ifthenelse{#4}%
	{%
	#5%
	\addtocounter{#2}{#1}%
	\forloop[#1]{#2}{\value{#2}}{#4}{#5}%
	}%
	{%
	}%
}%

\newcommand{\ctbd}[1]{}

\newcommand{\lc}{light curve}
\newcommand{\lcs}{light curves}

\newcommand{\teff}{\ensuremath{T_{\rm eff}}}
\newcommand{\logg}{\ensuremath{\log{g}}}

\newcommand{\feh}{\ensuremath{\rm [Fe/H]}}

\newcommand{\rsun}{\ensuremath{R_\odot}}

\newcommand{\rstar}{\ensuremath{R_\star}}

\newcommand{\rearth}{\ensuremath{R_\oplus}}

\newcommand{\refsec}[1]{\mbox{\S\ \ref{sec:#1}}}

\newcommand{\tbn}[1]{\tablenotemark{#1}}

%% file: anomaly.tex
1002.5198 & 1002.7241 & 2 & EXCLUDE \tablenotemark{a}\\
1014.5146 & 1016.7214 & 2 & SAFE MODE \tablenotemark{b}\\
1033.2932 & 1033.3341 & 2 & TWEAK \tablenotemark{c}\\
1056.4853 & 1056.8313 & 2 & COARSE POINT \tablenotemark{d}\\
1063.2896 & 1064.3726 & 2 & EARTH POINT \tablenotemark{e}\\
1073.3632 & 1073.3632 & 2 & ARGABRIGHTENING \tablenotemark{f}\\
1079.1866 & 1079.1866 & 2 & TWEAK \\
1080.8008 &  1080.8212& 2 & ARGABRIGHTENING \\
1088.4018 & 1089.3213 & 2 & COARSE POINT \\
1093.2239 & 1093.2239 & 3 & EARTH POINT \\
1100.4163 & 1100.5389 & 3 & COARSE POINT \\
1104.5233 & 1104.5438 & 3 & COARSE POINT \\
1113.4116 & 1117.3346 & 3 & MEMORY ERROR \tablenotemark{g}\\
1123.5461 & 1124.4248 & 3 & EARTH POINT \\
1206.2584 & 1206.2584 & 4 & TWEAK \\
1216.4137 & 1217.3128 & 4 & EARTH POINT \\
1229.8386 & 1233.8231 & 4 & SAFE MODE \\
1307.9995 & 1309.2869 & 5 & -  \\
1336.7713 & 1337.6091 & 5 & -  \\
1399.5042 & 1400.3828 & 6 & EARTH POINT \\
1431.2374 & 1432.1978 & 6 & EARTH POINT 

%% file: KOI_short.tex
1.01 & 11446443 & 2.470613 & - & 955.76257 & - & - & - & - & - & sod \tablenotemark{a}\\ 
2.01 &  10666592  & 2.204735 &  2.204760  & 954.35780 & 965.37828  &  6.10e-03  &  7.55e-02  &  206.05  &  179.41  & \\ 
3.01 &  10748390  & 4.887800 &  4.887970  & 957.81254 &  967.579324  &  2.60e-03  &  2.39e-02  &  439.78  &  55.43  & \\ 
157.04 &   6541920   &  46.687100  &  22.689217  &  -  &  981.433477  &  6.00e-04  &  1.11e-02  &  58.68  &  37.10  &  wp \tablenotemark{b}\\ 
157.05 &   6541920   &  118.363800  &  113.449430  &  -  &  1026.812678  &  8.00e-04  &  2.10e-03  &  15.36  &  21.01  &  wp
 \\ 
375.01 &  12356617  &  600.000000  &  -  &  1072.22382  &  -  &  -  &  -  &  -  &  -  &  snd \tablenotemark{c}\\
1099.01 &  2853093  &  161.526600  &  -  &  1031.00152  &  -  &  -  &  -  &  32.38  &  5.94  &  ld \tablenotemark{d} \\
1448.01 &  9705459  &  2.486600  &  2.486616  &  967.10929  &  964.62009  &  4.25e-02  &  4.46e-02  &  167.96  &  135.49  &  lr \tablenotemark {e}\\
1888.01 &  10063802  &  120.019000  &  -  &  967.18256  &  -  &  -  &  -  &  10.05  &  13.06  &  ls \tablenotemark{f} \\
2066.01 &  3239671  &  147.972400  &  -  &  1096.09006  &  -  &  -  &  -  &  -1.00  &  -1.00  &  nd \tablenotemark{g} \\
... & ... &  ... & ... & ... & ... & ... & ... & ... & ... & ... \\

%% file: fp_short.tex
2442359 & 13.931 & 19 24 53.566 & +37 46 53.58 & 965.8428 & 0.55283 & -0.0025 & - & 26.16 \\ 
3336765 & 13.617  & 19 19 49.397 & +38 27 49.07 & 965.1613 & 1.84483 & - & -0.00032 & 38.93 \\
3852258 & 13.819 & 19 27 55.433 & +38 58 16.25 & 968.1106 & 5.75838 & - & 0.005 & 17.48 \\ 
4035667 & 10.035 & 18 58 29.566 & +39 10 56.68 & 965.5837 & 2.87366 & -0.007 & 0.005 & -1.00 \\ 
4072333 & 15.664  & 19 41 19.236 & +39 09 45.14 & 965.5638 & 2.00775 & - & 0.0030 & 37.69 \\
4077901 & 13.007 & 19 45 42.948 & +39 08 01.54 & 967.7249 & 6.05444 & + & 0.002 & 27.22 \\ 
4270565 & 15.154 & 19 34 24.946 & +39 23 18.96 & 972.2024 & 18.01143 & 0.007 & -0.007 & 28.47 \\ 
5443775 & 12.936  & 19 21 20.671 & +40 39 11.02 & 967.8765 & 3.30754 & - & -0.00034 & 73.09 \\  
5565497\tbn{c}& 11.521 & 19 57 31.702 & +40 45 29.30 & 966.11575 & 1.412532 & -0.0126 & -0.0011 & 131.4 \\
5622812 & 12.832 & 19 30 41.270 & +40 52 10.34 & 964.8384 & 0.10304 & -0.005 & -0.005 & 49.76 \\ 
5649325\tbn{c} & 13.560 & 19 55 55.289 & +40 52 19.34 & 1196.35 & 195.63 & 0.0016 & - & 7.81\\
... & ... & ...& ... & ...& ... & ... & ... & ... 

%% file: stellar_short.tex
2985587 & 13.910 & 6023 & 4.278 & -0.069 & 1.276 &  0.34 & 0.29 \\ 
3128552 & 14.523 & 5530 & 4.673 & -0.248 & 0.761 &  0.41 & 0.25 \\ 
3240049 & 11.557 & 4435 & 2.127 & -0.065 & 17.188 &  0.65 & 0.09 \\ 
3245969 & 15.681 & 4825 & 4.789 & 0.054 & 0.594 &  0.59 & 0.13 \\ 
3328026 & 15.147 & 5681 & 4.508 & 0.147 & 0.947 &  0.42 & 0.24 \\ 
3345675 & 15.635 & 4105 & 4.628 & 0.137 & 0.598 &  0.57 & 0.17 \\ 
3346154 & 14.575 & 5513 & 4.490 & 0.118 & 0.957 &  0.45 & 0.22 \\ 
3439096 & 13.799 & 4940 & 3.119 & -0.340 & 5.493 &  0.52 & 0.18 \\ 
3541946 & 13.597 & 5537 & 4.728 & -0.140 & 0.711 &  0.43 & 0.24 \\ 
3558849 & 14.218 & 5938 & 4.432 & -0.410 & 1.052 &  0.33 & 0.30 \\ 
3728432 & 15.646 & 4371 & 4.583 & -0.254 & 0.696 &  0.59 & 0.13 \\ 
3764879 & 14.024 & 5845 & 4.641 & 0.006 & 0.809 &  0.38 & 0.27 \\ 
3834322 & 15.397 & 4627 & 4.631 & -0.004 & 0.698 &  0.62 & 0.11 \\
4150804\tablenotemark{d} & 12.888 &  - & - & - & - & 0.40 & 0.25 \\
...     & ...    & ...  & ...   & ...   & ...  & ... & ...

%% file: planet_short.tex
2985587 & 1159.6256 & 1.5 & 3.375816 & 0.52 & 1.64040 & 0.01178 & 0.84 & $0.68 \pm 0.19$ & $42.99 \pm 2.9$ & 40.83 & 14.78 & 1.44 & \\ 
3128552 & 1219.6102  & 1.2 & 2.504621 & 0.18 & 1.06137 & 0.01278 & 0.90 & $0.66\pm0.19$ & $24.26\pm0.86$ & 27.167 & 9.826& 4.16 & KOI2055 \tbn{b},O12 \tbn{c}\\
3128552 & 1298.5430 & 4.9 & 4.025928 & 0.75 & 1.1627 & 0.0140 & 1.1 & $0.68 \pm 0.18$ & $16.91 \pm 0.50$ & 33.43 & 10.76 & 4.16 & KOI2055,O12\\ 
3218908 & 1187.6632 & 1.2 & 4.152573 & 0.54 &0.9736 & 0.0128 & 0.96 & $0.64\pm0.20$ & $18.73\pm0.61$ & 27.1020 & 12.7280 & 4.09 & KOI1108,O12 \\
3240049 & 1234.98694 & 0.43 & 2.9454270 & 0.094 & 61.9190 & 0.03301 & 0.99 & $0.95 \pm 0.037$ & $23.61 \pm 0.33$ & 24.94 & 34.13 & 0.62 & T12 \tbn{d}\\ 
3245969 & 1301.042 & 15 & 11.39060 & 1.8 & 1.8410 & 0.0284 & 1.9 & $0.46 \pm 0.38$ & $20.4 \pm 1.7$ & 36.80 & 10.72 & 18.97 & KOI1101,T12\\ 
3326377 & 1188.2283  & 1.8 & 198.7039  & 20 & 3.7018 & 0.0379 & 1.8 & $0.67\pm0.15$ & $6.127\pm0.080$ & 19.5130 & 24.9130 & 3.22 & KOI1830,O12 \\ 
3328026 & 1102.5054 & 1.7 & 2.115515 & 0.39 & 1.75795 & 0.01701 & 0.73 & $0.58 \pm 0.12$ & $10.6 \pm 1.1$ & 103.06 & 35.56 & 6.96 & T12\\ 
3345675  &1203.14403& 0.12 &120.002673 & 0.80 & 14.73987 & 0.22586 & 0.28 & $0.90\pm0.00$ & $26.70\pm0.000$& 8.03 & 17.837 & 18.34 & wp\tbn{e},PHt \tbn{f}\\
3346154 & 1231.72962 & 0.31 & 1.9527041 & 0.057 & 1.855891 & 0.01777 & 0.78  & $0.48\pm0.20$ & $53.7\pm1.1$ & 105.71 & 22.92 & 4.40 & T12\\
3439096 & 1108.8290 & 1.3 & 2.975892 & 0.21 & 7.61918 & 0.01271 & 0.50 & $0.53 \pm 0.17$ & $11.74 \pm 0.21$ & 32.91 & 23.49 & 3.71 & \\ 
3541946 & 1270.418 & 12 & 1.311837 & 0.13 & 0.9854 & 0.0127 & 5.2 & $0.17 \pm 0.83$ & $20.6 \pm 2.0$ & 61.05 & 37.19 & 1.50 & KOI624,O12\\ 
3558849 & 1112.9840 & 1.5 & -258 &-& 6.67780 & 0.05817 & 0.61 & $0.34\pm0.12$ & $2.919\pm0.013$ & - & - & 2.11 & \\
3728432 & 1314.79711 & 0.22 & 3.9086360 & 0.043 & 1.92017 & 0.02528 & 0.47 & $0.30 \pm 0.14$ & $27.83 \pm 0.20$ & 107.92 & 19.85 & 15.22 & T12\\ 
3764879 & 1118.75219 & 0.68 & 1.3069750 & 0.067 & 1.3331 & 0.0151 & 1.0 & $0.73 \pm 0.17$ & $49.29 \pm 1.80$ & 112.23 & 31.15 & 1.69 & T12\\ 
3834322 & 1087.18555 & 0.12 & 0.49844300 & 0.0040 & 0.99103 & 0.01301 & 0.47 & $0.46 \pm 0.16$ & $50.41 \pm 0.68$ & 17.46 & 7.65 & 8.90 & T12\\ 
4150804 &1271.38683 & 0.69 & 160.8818 & 10&  - & 0.07985 & 0.64 & $0.365\pm0.080$ & $4.801\pm0.013 $ & 12.06 & 24.85 & 0.35 & PH\tbn{f}\\
4245933 & 1165.1932 & 1.9 & 11.25629 & 1.3 & 0.81285 & 0.01408 & 0.45 & $0.46 \pm 0.17$ & $3.23 \pm 0.027$ & 25.29 & 11.88 & 13.00 & T12\\ 
4271474 & 1262.46308 & 0.40 & 21.86183 & 1.1 & 1.71411 & 0.02580 & 0.59 & $0.35 \pm 0.14$ & $15.71 \pm 0.11$ & 20.09 & 9.26 & 16.18 & \\ 
4275117 & 1146.8945 & 5.1 & 2.013042 & 0.12 & 1.2924 & 0.0139 & 1.0 & $0.62 \pm 0.18$ & $43.66 \pm 3.33$ & 34.05 & 11.09 & 3.64 & T12\\ 
4552729 & 1427.64700 & 0.16 & 97.461490 & 0.62 & 4.174580 & 0.039153 & 1.1 & $0.47\pm0.14$ & $7.474\pm0.078$ & 44.69 & 10.07 & 3.68 & PH\\
4858610 & 1128.2926 & 1.0 & 2.722874 & 0.18 &1.17141  & 0.01202 & 0.61 & $0.57 \pm 0.18$ & $16.72 \pm 0.30$ & 32.15 & 12.00 & 5.18 & \\ 
4927315 & 1077.3419 & 1.5 & 11.76856 & 1.5 & 1.9340 & 0.0192 & 1.2 & $0.67 \pm 0.18$ & $24.2 \pm 1.1$ & 31.24 & 17.07 & 3.49 & T12\\ 
4951249 & 1072.5112 & 1.4 & 4.523406 & 0.24 & 1.29861 & 0.01574 & 0.62 & $0.55 \pm 0.18$ & $18.02 \pm 0.59$ & 40.99 & 11.01 & 6.36 & T12\\ 
5008501 & 1073.17351 & 0.38 & 0.9680371 & 0.023 & 1.00658 & 0.01290 & 0.63  & $0.65\pm0.18$ & $55.4\pm1.3$ & 34.65 & 17.00 & 1.83 & PH,T12\\
5023843 & 1320.7241 & 1.2 & -358 &-& 6.1223 & 0.0543 & 1.5 & $0.3\pm1.0$ & $11.00\pm0.15$ & - & - & 2.83 & PH \\
5128673  &1353.1823 & 5.2 & 87.9654 & 36 & 2.25057 & 0.02746 & 0.44 &  $0.1435\pm0.0049$ & $3.368\pm0.048$ & 28.37 & 8.66 & 2.81 & blend\tbn{g}\\
...      & ...      & ... & ...     & ... & ...    & ...     & ... & ... & ... & ... & ... & ... & ...  